\documentclass[useAMS,usenatbib,usegraphicx,letterpaper]{mn2e}

\usepackage{multirow} 
\usepackage{epsfig}
\usepackage{subfigure}

\newcommand{\etal}{et al.}

\title[IRAS~13197--1627 with {\it XMM--Newton}] {IRAS~13197--1627 has
  them all: Compton--thin absorption, photo--ionized gas, thermal
  plasmas, and a broad Fe line} \author[G.\ Miniutti \etal]
{G.~Miniutti$^1$\thanks{miniutti@ast.cam.ac.uk}, G.  Ponti$^{2,3}$,
  M.~Dadina$^3$, M.~Cappi$^3$ and G.~Malaguti$^3$\\
  $^1$Institute of Astronomy, Madingley Road, Cambridge CB3 0HA, UK \\
  $^2$ Dipartimento di Astronomia dell' Universit\'a degli Studi di
  Bologna, via Ranzani 1, I--40127, Bologna, Italy\\
  $^3$ IASF/INAF Sezione di Bologna, via Gobetti 101, I--40129,
  Bologna, Italy }

\pagerange{\pageref{firstpage}--\pageref{lastpage}}
\pubyear{2001}

\usepackage{times}

\begin{document}

\label{firstpage}

 \maketitle

\begin{abstract}
  We report results from the {\it XMM--Newton} observation of
  IRAS~13197--1627, a luminous IR galaxy with a Seyfert 1.8 nucleus.
  The hard X--ray spectrum is steep ($\Gamma\sim 2.5$) and is absorbed
  by Compton--thin (N$_{\rm{H}} \sim 4\times 10^{23}$~cm$^{-2}$)
  neutral gas. We detect an Fe K$\alpha$ emission line at 6.4~keV,
  consistent with transmission through the absorber.  The most
  striking result of our spectral analysis is the detection of a
  dominant X--ray reflection component and broad Fe line from the
  inner accretion disc. The reflection--dominated hard X--ray spectrum
  is confirmed by the strong Compton hump seen in a previous {\it
    BeppoSAX} observation and could be the sign that most of the
  primary X--rays are radiated from a compact corona (or base of the
  jet) within a few gravitational radii from the black hole. We also
  detect a relatively strong absorption line at 6.81~keV which, if
  interpreted as Fe\textsc{xxv} resonant absorption intrinsic to the
  source, implies an outflow with velocity $\sim 5\times
  10^3$~km~s$^{-1}$. In the soft energy band, the high--resolution RGS
  and the CCD--resolution data show the presence of both
  photo--ionized gas and thermal plasma emission, the latter being
  most likely associated with a recent starburst of
  15--20$M_\odot$~yr$^{-1}$.
\end{abstract}


\section{Introduction}

IRAS~13197--1627 (also MCG--03-34-064, z=0.01654) is a ``warm''
($f_{\rm{25\mu m}}/f_{\rm{60\mu m}} \sim 0.48$) and luminous
IR--galaxy with L$_{\rm{IR}}$=L(8--1000~$\mu m$)=$1.7\times 10^{11}
L_\odot$ (Sanders et al. 2003). It was first classified as a
moderately reddened Seyfert~2 by Osterbrock and de Robertis (1985) \&
de Robertis, Hutchings \& Pitts (1998).  However, as pointed out by
these authors, the source has peculiar emission--line properties: in
particular its H$\alpha$, [O$_{\rm I}$], [N$_{\rm II}$], and [S$_{\rm
  II}$] are both exceptionally broad for Seyfert~2s (with FWHMs from
480~km~s$^{-1}$ to 860~km~s$^{-1}$) and asymmetric, with excess flux
blue--wards of their centres.  In the UV, most of the lines are broad at
the level seen in the Optical, and C$_{\rm IV}$ as a
width comparable to those typically measured in Seyfert~1 galaxies.
Aguero et al (1994) were able to disentangle broad and narrow
components of the H$\alpha$ and H$\beta$ lines which led to a Seyfert~1.8
classification. A broad component to H$\alpha$ was also detected
by Young et al (1996) in polarised light. Signatures of the presence
of Wolf--Rayet stars have been reported by Cid Fernandes et al (2004)
who also estimate that about 25 per cent of the stellar population is
relatively young (less than 25~Myr). Extended Radio emission is
also detected with a linear extension of about 280~pc, almost aligned
with the host galaxy major axis (Schmitt et al. 2001a; 2001b) while
the 1.4~GHz luminosity is $1.6\times 10^{30}$~erg~s$^{-1}$~Hz$^{-1}$
(Condon et al. 1996).

In the X--rays, the source was first observed with {\it ASCA} in 1995
revealing a very steep spectral shape (photon index $\Gamma \simeq
3$), a large absorbing column density $N_{\rm{H}} \simeq 6\div 8
\times 10^{23}$~cm$^{-2}$, a narrow Fe emission line at 6.4~keV, and a
``soft excess'' component emerging below about 2~keV (Ueno 1997). A
{\it BeppoSAX} observation was obtained in 1998 and was first
published by Risaliti (2002) in a statistical study of 20 bright
Seyfert~2 galaxies. A variation in the absorbing column density
between the {\it ASCA} and the {\it BeppoSAX} observations was later
suggested by Risaliti, Elvis \& Nicastro (2002). A more detailed
analysis of the {\it BeppoSAX} data was presented by Dadina \& Cappi
(2004). The absorbing column was found to be dependent on the adopted
spectral model so that the column density variation could not be
confirmed (we shall discuss this point in our analysis below). The
better quality, broadband {\it BeppoSAX} data also revealed spectral
complexity that could be modelled in terms of a partial
covering scenario or of a reflection--dominated model in
which X--ray reflection from the accretion disc dominates the hard
spectral shape. In both cases, Dadina \& Cappi estimated an intrinsic
2--10~keV luminosity of $\sim 2 \times 10^{44}$~erg~s$^{-1}$, making
IRAS~13197--1627 the nearest and brightest type--1.8 quasar known to
date. An absorption line at $\sim 7.5$~keV was also detected and, if
interpreted as He--like Fe resonant absorption, the line energy
implies an outflow with a velocity of the order of 0.1~c.

Here we report on results from a new {\it XMM--Newton} observation of
IRAS~13197--1627 and we present detailed spectral analysis at both CCD
and gratings resolution allowing us to better understand the nature of
the hard and soft X--ray emission components in this luminous IR
galaxy. 

\begin{figure}
\begin{center}
  \includegraphics[width=0.385\textwidth,height=0.455\textwidth,angle=-90]{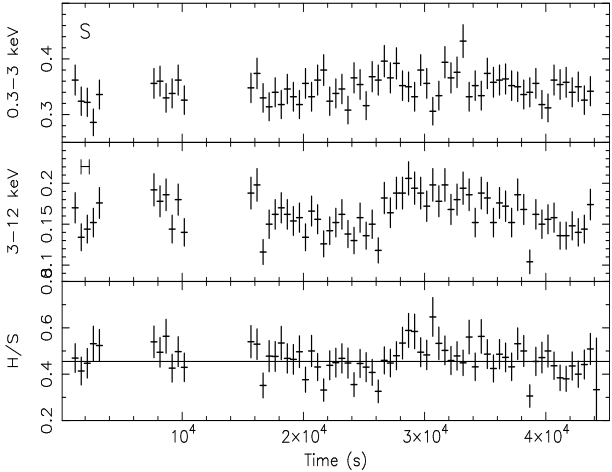}
\end{center}
\caption{We show a soft (S: 0.3--3~keV) and a hard (H: 3--12~keV)
  EPIC--pn source light curve together with the resulting hardness
  ratio (bottom panel). As revealed by fitting a constant to the data,
  the hard band is more variable than the soft one ($\chi^2_r=1.65$ vs.
  $\chi^2_r=0.85$ for 67 dof). However, the hardness ratio is still
  consistent with a constant value throughout the observation.}
\end{figure}

\section{The XMM--Newton observation and first--look analysis}

IRAS~13197--1627 was observed by {\it XMM--Newton} on 2005 January 24
for a total exposure of 45~ks. The data have been processed starting
from the {\small ODF} files with the {\it XMM--Newton} {\small SAS}
software (version 6.5.0). Source spectra and light curves of the EPIC
cameras were extracted from circular regions centred on the source,
while background products were extracted from off--set regions close
to the source. We remind here that the background region of the pn
camera has to be taken as close to the centre as possible to avoid
contamination from the strong spatially dependent fluorescent Cu
K$\alpha$ emission line originating from the electronics board mounted
below the pn CCD which would result in a spurious absorption line
around 8~keV in the (background subtracted) source spectrum. With our
choice of the background extraction region, such contamination is
negligible. After filtering for periods of high background the net
exposure is 37~ks in the pn camera and 39~ks (43~ks) in the MOS 1 (MOS
2) detectors. The EPIC spectra were grouped to have at least 20 counts
per bin to ensure the validity of $\chi^2$ statistical analysis. The
RGS were operated in the standard spectroscopy mode and standard data
reduction was performed resulting in a net exposure of about 43~ks in
the RGS detectors.

\begin{figure}
\begin{center}
 \includegraphics[width=0.355\textwidth,height=0.455\textwidth,angle=-90]{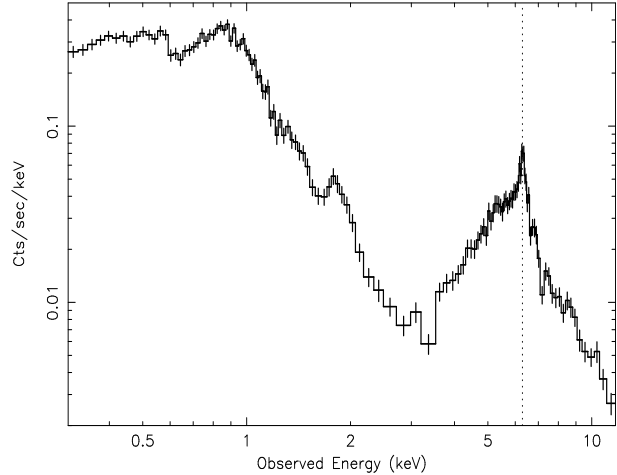}
\end{center}
\caption{The broadband (0.3--12~keV) EPIC--pn spectrum of
  IRAS~13197--1627. The spectrum exhibits strong curvature in the hard
  band which is a clear sign of absorption by a substantial column of
  gas in the line of sight. Fe emission is detected at 6.4~keV in
  the rest--frame (vertical dotted line) together with a deep
  absorption edge at 7~keV. A soft excess is present below
  $\sim$~3~keV and is characterised by a bump around 0.9~keV. Data
  have been rebinned for visual clarity.}
\end{figure}

In Fig.~1 we show the background--subtracted source light curve in a
soft (S: 0.3--3~keV) and a hard (H: 3--12~keV) energy band after
removal of residual background flares. No strong variability is seen
in the soft band and a fit with a constant is very satisfactory
($\chi^2_r\simeq 0.85$ for 67 dof). More fluctuations are present in
the hard band ($\chi^2_r=1.65$). In the bottom panel, we plot the
hardness ratio and compare it with the best--fitting constant value
producing an acceptable fit with $\chi^2_r=1.1$. Given the limited
evidence for spectral variability, in the following we consider the
time--averaged spectrum of the source only. The broadband EPIC--pn
spectrum of IRAS~13197--1627 is shown in Fig.~2. The MOS data are not
shown for visual clarity and agree very well with the pn.  The main
features of the spectrum are i) strong curvature in the hard band with
a $\sim$~3--4~keV low--energy cut--off and a deep spectral drop around
7~keV where the Fe absorption edge is expected; ii) a clear narrow Fe
emission feature at 6.4~keV (vertical dotted line in Fig.~2); iii) a
soft excess emerging below 2--3~keV which seems structured rather than
smooth, especially around 0.9--1~keV. In this paper we adopt standard
cosmology parameters ($H_0=70$~km~s$^{-1}$, $\Lambda_0=0.73$, and
$q_0=0$).

\begin{figure}
\begin{center}
  \includegraphics[width=0.35\textwidth,height=0.455\textwidth,angle=-90]{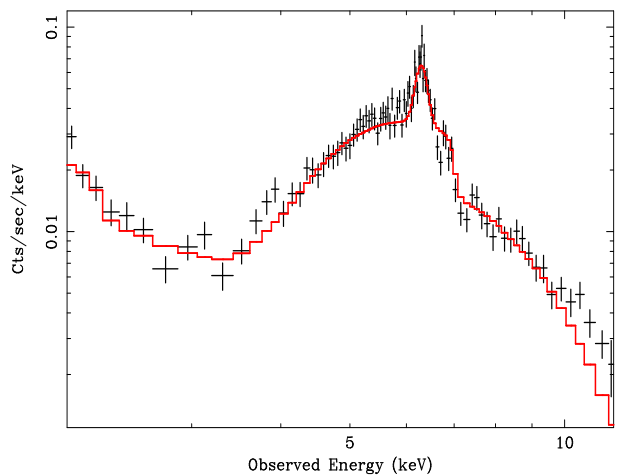}
\end{center}
\caption{The 2--12~keV pn spectrum is shown together with a simple
  model (solid line) comprising a power law absorbed by a column
  density of $\sim 6.2\times 10^{23}$~cm$^{-2}$ of neutral matter at
  the redshift of the source and a resolved ($\sigma \sim 90$~eV)
  Gaussian emission line at 6.4~keV. Only the pn data are shown but
  all spectral analysis is performed simultaneously with the MOS data
  as well. Notice several residuals: absorption structures are seen
  around 6.8~keV and 7.1~keV.  Moreover the model systematically
  underestimates the data in the 5--6~keV band and above 10~keV. In
  the Figure, data have been rebinned for visual clarity.}
\end{figure}

\section{The 2--12~keV spectrum of IRAS~13197--1627}

In our analysis, we consider joint fits to the EPIC--pn and MOS (1 and
2) cameras. The pn data are considered up to 12~keV, while for the MOS
we limit our analysis up to 9.5~keV due to the lack of
signal--to--noise above that energy. As a first attempt to describe
the hard spectrum above 2~keV, we consider a simple model in which a
power law is absorbed by neutral matter at the redshift of the source
(the {\small ZWABS} model in {\small{XSPEC}} ). We also include a
second power law component emerging at soft energies to account for
the soft excess. The two power law slopes ($\Gamma_h$ and $\Gamma_s$
respectively) are allowed to be different.  The observed Fe line at
6.4~keV is modelled with a Gaussian emission line with width, energy,
and normalisation free to vary. All subsequent fits include Galactic
absorption with column density fixed at its nominal value ($5.8\times
10^{20}$~cm$^{-2}$, Dickey \& Lockman 1990).

With this simple model, we obtain an acceptable fit to the 2--12~keV
data with $\chi^2=519$ for 447 degrees of freedom (dof). The hard
power law has a slope $\Gamma_h=2.9\pm 0.4$, marginally consistent
with the soft one ($\Gamma_s=2.4\pm 0.4$). The ratio of the soft to
the hard power law normalisations is of about 1--2 per cent only.  The
neutral absorber affects the hard band above $\sim$3~keV and is
Compton--thin with a column density of $6.2\pm 0.5 \times
10^{23}$~cm$^{-2}$.

The Fe emission line is clearly detected ($\Delta\chi^2=160$ for 2
more free parameters) and has an energy of $6.40\pm 0.02$~keV
indicating an origin in neutral matter.  The line has an observed
equivalent width (EW) of $140\pm 20$~eV when computed with respect to
the unabsorbed continuum, to be compared with the theoretical
prediction of 80~eV for a line transmitted through the observed column
density (e.g. Matt 2002).
\begin{figure}
\begin{center}
 \includegraphics[width=0.35\textwidth,height=0.455\textwidth,angle=-90]{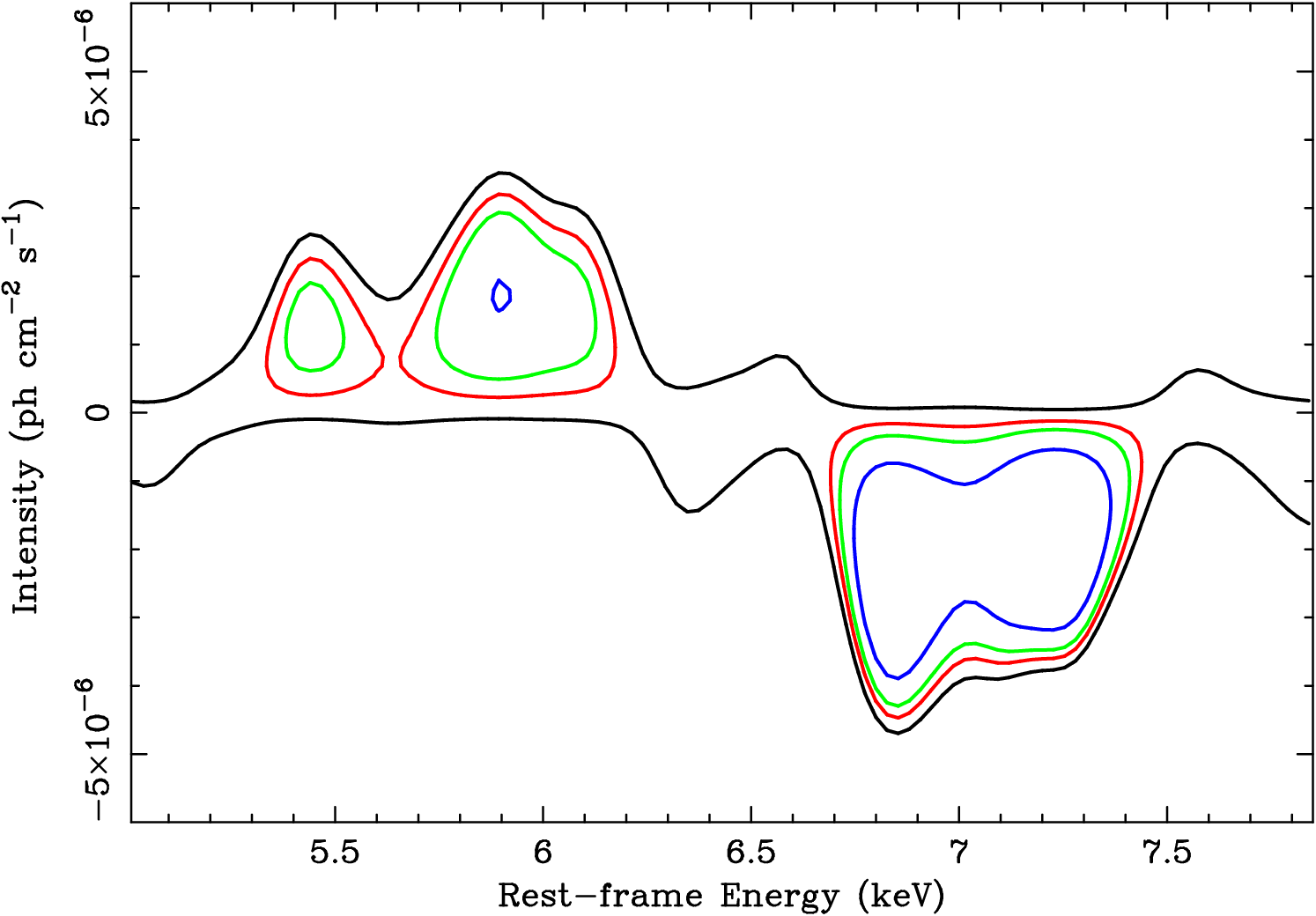}
{\vspace{0.2cm}}
 \includegraphics[width=0.35\textwidth,height=0.455\textwidth,angle=-90]{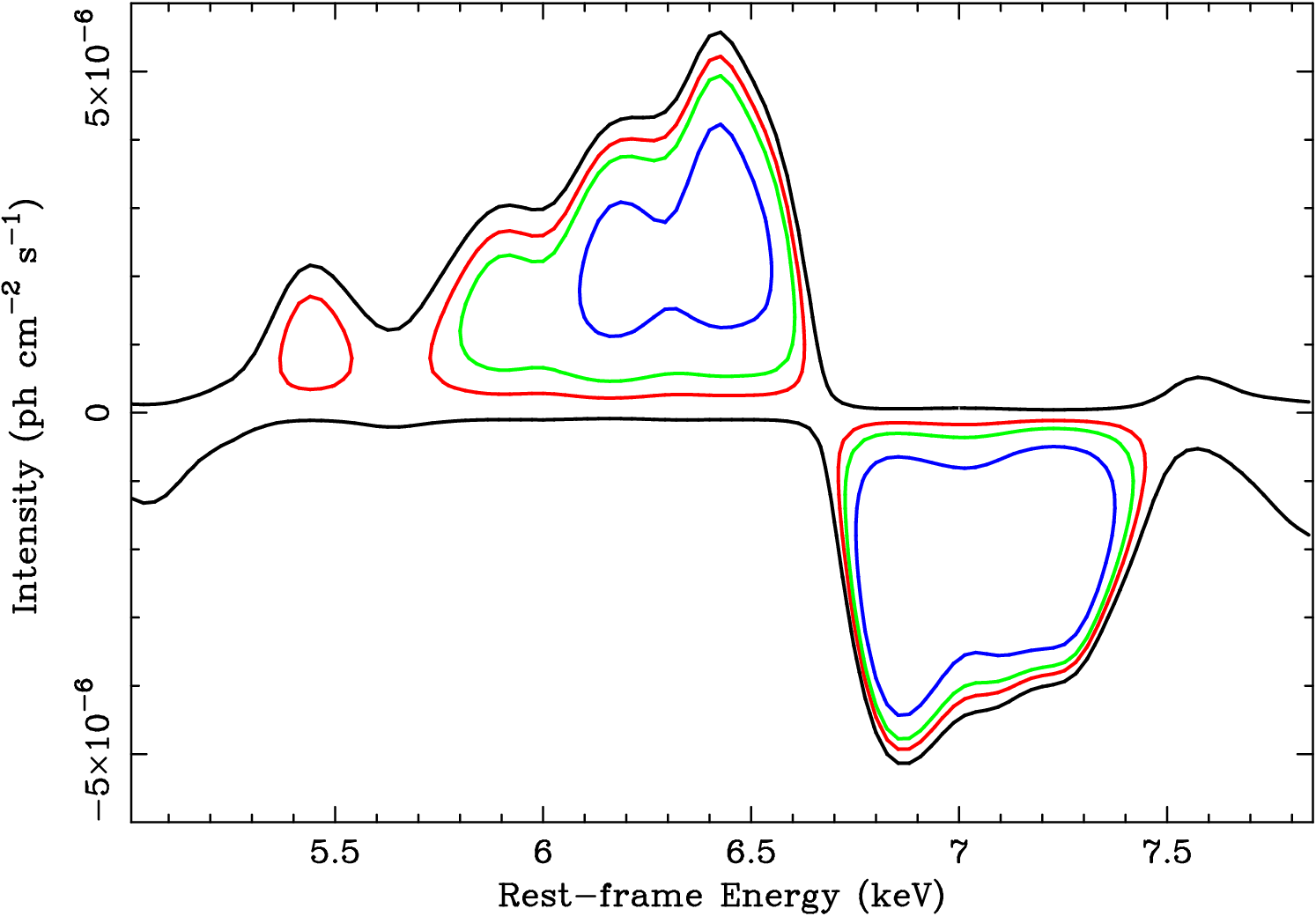}
\end{center}
\caption{{\bf{Top:}} A Gaussian line is added to the baseline model
  and its energy and normalisation are varied. The contours represent
  an improvement of $\Delta\chi^2=-1, -2.3, -4.61, -9.21$ in fitting
  the joint pn--MOS data. The $\Delta\chi^2=-1$ contour (outermost) is
  shown as a reference of the best--fitting continuum model.
  {\bf{Bottom:}} In this case the Fe K$\alpha$ line of the baseline
  model is forced to be unresolved. The shape of the residuals
  strongly suggests the presence of X--ray reflection from the disc
  (broad Fe emission line and broad Fe absorption edge).}
\end{figure}

We measure an Fe line width $\sigma=90^{+40}_{-20}$~eV, significantly
broader than the EPIC cameras spectral resolution ($\sim$40~eV in
$\sigma$). The measured Fe line width corresponds to a
FWHM$=1.21^{+0.44}_{-0.33}\times 10^4$~km/s.  Under the assumption
that the line emitting gas is gravitationally bound and that gas
motion occurs in randomly oriented circular orbits, the line width
places the emitting gas {\emph{within}} $2\times 10^3$ gravitational
radii ($r_g=GM/c^2$) from the central black hole (Krolik 2001). If the
line, as is likely, is transmitted through the Compton--thin absorber,
the line width would place the absorber much closer to the nucleus
than often thought (e.g. Galactic discs, Maiolino \& Rieke 1995; dust
lanes, Malkan, Gorijn \& Tam 1998; Guainazzi, Matt \& Perola 2005;
starburst clouds, Weaver 2001; see also Lamastra, Perola \& Matt
2006).  However, a much closer location of the absorbing gas,
consistent with the measured line width, has been proposed by e.g.
Elvis (2000; 2004). The case for an absorber located close to the
nucleus is particularly compelling in NCG~1365 (Risaliti et al.  2005)
in which rapid transitions between Compton--thin and Compton--thick
absorption have been observed. On the other hand, as we shall discuss
below, the observed Fe line width could be due to an unmodelled
broader component to the Fe line profile.

Although our simple spectral model (hereafter the ``baseline model'')
provides a reasonable description of the hard spectrum, several
residuals are present as shown in Fig.~3.  Two absorption features are
seen around 6.8~keV and 7.1~keV, while positive residuals are left at
high energy (above 10~keV) and in the 5--6~keV band where the model
seems to systematically underestimate the data. As a first check for
the significance of the residuals, we add an unresolved Gaussian line
to the baseline model, vary its energy in the 4--8~keV range, its
normalisation in the $\pm1\times10^{-5}$~ph~cm$^{-1}$~s$^{-1}$ range,
and record the $\chi^2$ improvement (see Miniutti \& Fabian 2006 for
more details).  The results are presented in the top panel of Fig.~4
where we show the 68, 90, and 99 per cent confidence contour levels
for the additional Gaussian line rest-frame energy and intensity in
the most relevant band around the Fe K complex. The outermost line
shows the $\Delta\chi^2=-1$ contour as reference.

\begin{figure}
\begin{center}
 \includegraphics[width=0.35\textwidth,height=0.455\textwidth,angle=-90]{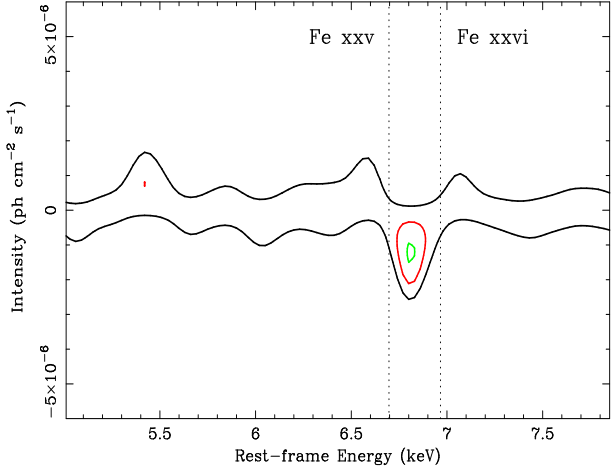}
\end{center}
\caption{Same as in Fig.~4, but now for the best--fitting model
  comprising Compton--thin absorption and reflection from the
  accretion disc, both with slightly super--solar Fe abundance (see
  text for details). The only remaining residual is a narrow
  absorption line at 6.8~keV. The rest--frame energies of the closest
  transitions (Fe\textsc{xxv} and Fe\textsc{xxvi} resonant absorption)
  are shown as vertical lines.}
\end{figure}

Although the contours should not be taken as a precise significance
test, they indicate the relative importance of the residuals. Fig.~4
clearly reveals both positive and negative residuals on the red and
blue side respectively of the Fe K$\alpha$ line (6.4~keV). If the Fe
K$\alpha$ emission line in the best--fitting model is forced to be
narrow (i.e. unresolved, as in most Compton--thin sources), the bottom
panel figure is obtained, where the positive residuals are very
reminiscent of X--ray reflection from the accretion disc resulting in
a skewed and broad Fe emission line and Fe absorption edge. With the
caveats mentioned above, the significance of the whole structure seem
to be higher than 99 per cent. A more detailed spectral modelling of
these absorption and emission features is presented below.

\subsection{X--ray reflection from the accretion disc and
  Fe\textsc{xxv} resonant absorption}

\begin{table}
\begin{center}
  \caption{The best--fitting parameters from joint fits to the
    2--12~keV (pn) and 2--9.5~keV (MOS) spectrum of IRAS~13197--1627
    ($\chi^2=442$ for 438 dof). 
}
\begin{tabular}{lcr}          
\hline
{\bf{Parameter}} & {\bf{Value}} & {\bf{F--test}} \\      
\hline
 \multicolumn{2}{l}{\bf{Continuum}} &  \\
\hline
$\Gamma_h$ & $2.5\pm 0.3$& \\
$N_h$~[$10^{-3}$~ph~cm$^{-2}$~s$^{-1}$]&  $5.7^{+1.6}_{-1.9}$& \\
$\Gamma_s$ & $2.1\pm 0.4$& \\
$N_s$~[$10^{-5}$~ph~cm$^{-2}$~s$^{-1}$]& $8.2 \pm 4.8$&  \\
\hline
 \multicolumn{2}{l}{\bf{Absorption}} &  $>99.99\%$\\
\hline
$N_H$~[$10^{23}$~cm$^{-2}$]& $3.9\pm 0.4$ & \\
\hline
 \multicolumn{2}{l}{\bf{Fe abundance}} &  $99.8\%$\\
\hline
$A_{Fe}$~[solar] & $1.5\pm 0.3$& \\
\hline
 \multicolumn{2}{l}{\bf{Transmitted Fe K$\alpha$ line} } & $>99.99\%$ \\
\hline
$E$~[keV]& $6.40\pm 0.02$ & \\
$\sigma$~[eV] & $< 85$& \\
$EW$~[eV] & $105\pm 18$& \\
\hline
 \multicolumn{2}{l}{{\bf Disc reflection} } &  $99.8\%$\\
\hline
$\xi$~[erg~cm~s$^{-1}$]& $<60$ & \\
$E_{Ni}$~[keV]& $7.5\pm 0.06$ & $94.8\%$ \\
$EW_{Ni}$~[eV] & $60\pm 40$& \\
$R$& $10\pm 3$& \\
$r_{in}$~[$r_g=GM/c^2$]& $7^{+7}_{-5}$& $99.1\%$ \\
$i$~[degrees]& $27\pm 17$& \\
\hline
 \multicolumn{2}{l}{\bf{Fe\textsc{xxv} absorption line}} & $98.1\%$ \\
\hline
$E$~[keV]& $6.81^{+0.08}_{-0.06}$ & \\
$EW$~[eV] & $-50\pm 30$& \\
\hline
 \multicolumn{2}{l}{\bf{Flux and Luminosity}} & \\
\hline
$F_{2-10}$~[$10^{-12}$~erg~cm$^{-2}$~s$^{-1}$] & $2.1\pm 0.2$& \\
$L^{\rm{obs}}_{2-10}$~[$10^{43}$~erg~s$^{-1}$] & $0.7\pm 0.2$& \\
$L^{\rm{est}}_{2-10}$~[$10^{43}$~erg~s$^{-1}$] & $4.2^{+2.8}_{-2.2}$& \\
\hline
\end{tabular}
\end{center}
The EW of the Fe K$\alpha$ emission
line is computed with respect to the unabsorbed continuum assuming
it is transmitted and not absorbed. If the line is assumed to be
absorbed by the same column as the continuum its EW is $160\pm
70$~eV. The F--test results are obtained by removing the relevant
component from the model and by re--fitting the data.  The F--test 
corresponding to $r_{in}$ in the {\emph{Disc reflection}} section is
relative to the relativistic blurring kernel, i.e. it 
provides the significance of the relativistic effects.
\end{table}

The shape of the residuals in Fig.~4 strongly suggests that both the
positive and negative residuals could be accounted for by a reflection
component from the accretion disc. We then add a relativistically
blurred reflection spectrum in which reflection continuum and emission
lines are computed self--consistently (from Ross \& Fabian 2005). The
model free parameters are the ionization parameter $\xi$, the
illuminating power law slope, the normalisation, and the Fe abundance.
To test whether the Fe abundance is different from solar, we let it
free to vary, replace the Compton--thin absorber model with one which
allows to vary the Fe abundance ({\tt ZVFEABS}) and forced the Fe
abundance in the two models to be the same. The illuminating power law
slope is forced to have the same photon index as the hard power law
continuum. The relativistic blurring is obtained by using a {\tt LAOR}
kernel in which the emissivity profile is fixed at its standard value
($\epsilon=r^{-3}$), the outer disc radius at $400~r_g$, while the
inner disc radius and observer inclination are free parameters. The
reflection spectrum is absorbed by the same column density as the
continuum.

We obtain a very significant improvement with respect to the previous
model, with $\Delta\chi^2=63$ for 5 more free parameters ($\chi^2=456$
for 442 dof). Since the reflection model does not include emission
from Ni (see Ross \& Fabian 2005), we add a Gaussian emission line
with energy between 7.4~keV and 7.6~keV and apply to it the same
relativistic blurring as for the reflection spectrum. We obtain a
marginal improvement of $\Delta\chi^2=6$ for 2 more free parameters.
Before discussing the best--fit parameters, we consider again the
procedure used to produce Fig.~4 on this best--fitting model to search
for any other possible residual (see Fig.~5). The
X--ray reflection from the disc accounted for all positive and
negative residuals (compare Fig.~5 with Fig.~4) except what appears to
be an absorption line at $\sim$~6.8~keV whose significance can be
roughly estimated to be between 90 and 99 per cent.

We then add a further narrow Gaussian absorption line to our spectral
model and find a marginal improvement of $\Delta\chi^2=8$ for 2 more
free parameters (98.1 per cent significance level according to the
F--test) for a line energy of $6.81^{+0.08}_{-0.06}$~keV with
equivalent width of $-50\pm 30$~eV and a final statistics of
$\chi^2=442$ for 438 dof. The absorption line energy suggests an
origin in resonant absorption due to Fe\textsc{xxv} or Fe\textsc{xxvi}
(see e.g.  Bianchi et al. 2005b) but is inconsistent with the expected
rest--frame energies (weighted mean of 6.697~keV and 6.966~keV
respectively). If Fe\textsc{xxv} is assumed, the line energy indicates
an outflow velocity of $5061^{+3600}_{-2700}$~km~s$^{-1}$.
 It should be stressed
that the inferred outflow velocity is remarkably similar to the
systemic velocity of the galaxy.  IRAS~13197--1627 is at redshift
z=0.016541 which corresponds to a receding velocity of
4959~km~s$^{-1}$. Such a coincidence raises the possibility that the
line has a local origin (i.e. Galactic) as it was recently suggested
by McKernan, Yaqoob \& Reynolds (2004; 2005) for the cases of extreme
relativistic outflows inferred from X--ray spectroscopy in several
AGNs (see e.g. Pounds et al 2003; Reeves, O'Brien, \& Ward 2003). 
\begin{figure}
\begin{center}
\includegraphics[width=0.35\textwidth,height=0.455\textwidth,angle=-90]{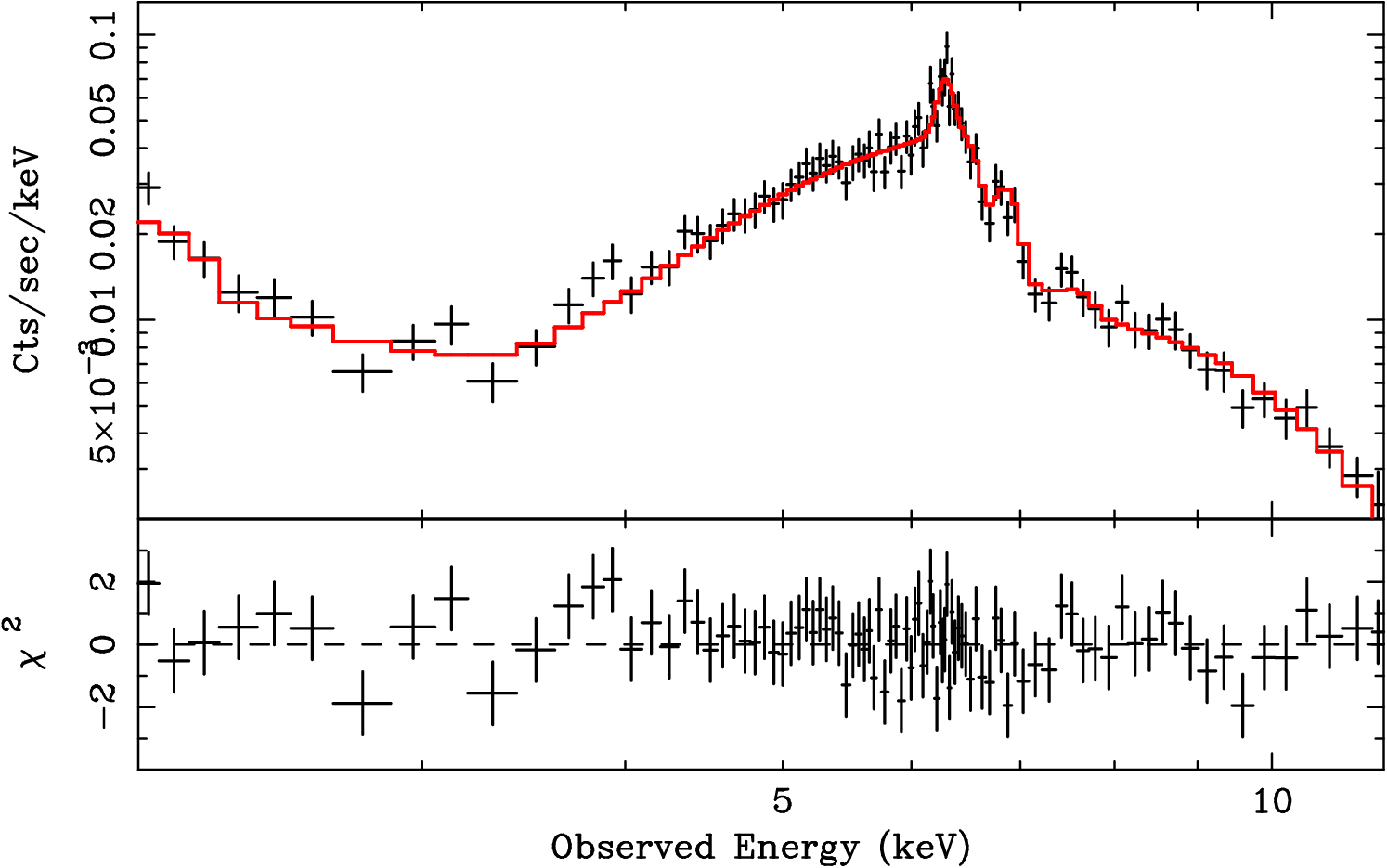}
{\vspace{0.2cm}}
\includegraphics[width=0.35\textwidth,height=0.455\textwidth,angle=-90]{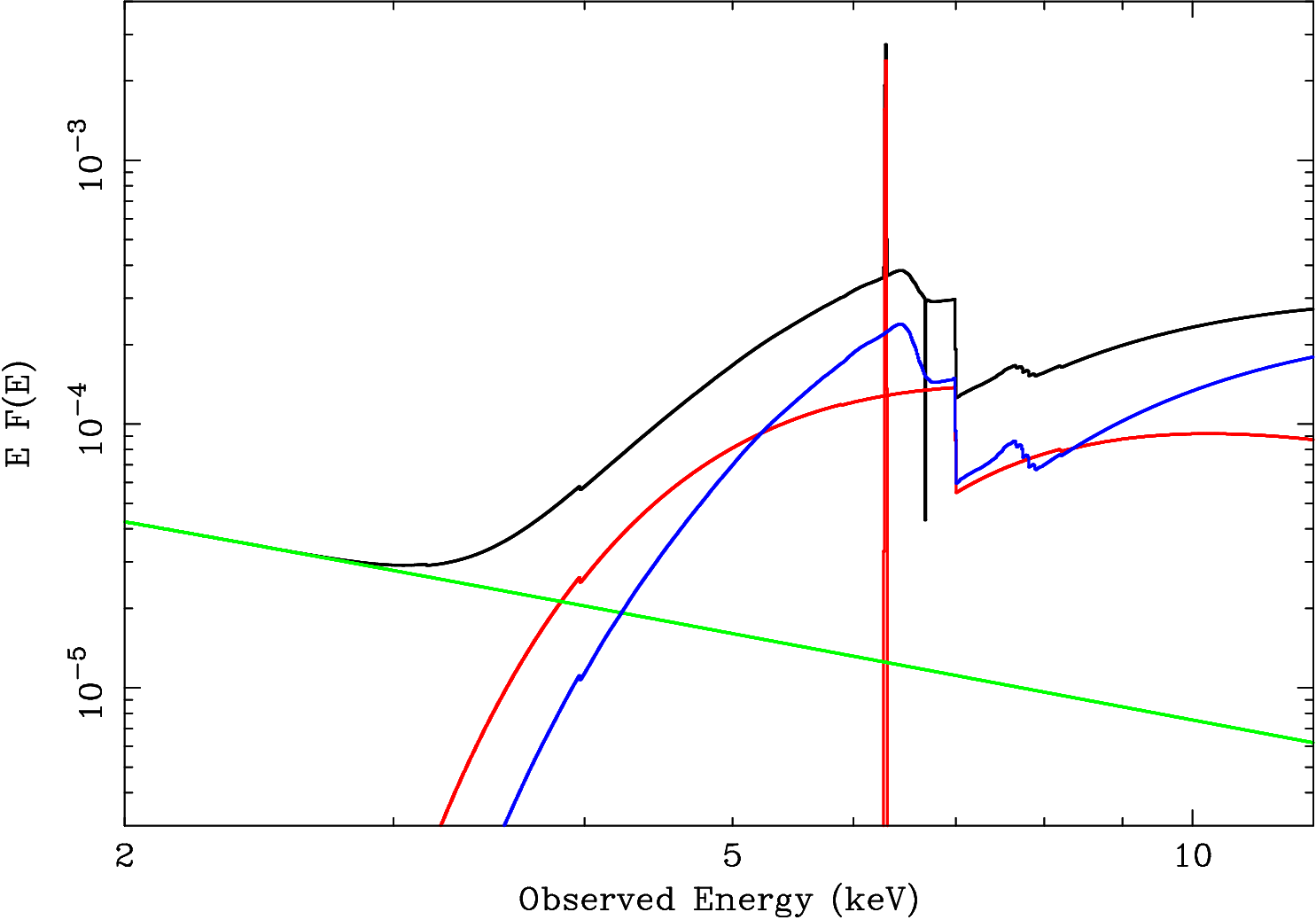}
\caption{In the top panel we show the result of our best--fitting
  model in the 2--12~keV band (see Table~1). Only the pn data are
  shown for clarity. In the bottom panel, we show the model
  components.}
\end{center}
\end{figure}


The data, residuals, and best--fitting model are shown in the two
panels of Fig.~6. We include the Ni emission line from the disc and
the 6.8~keV absorption line in the final model, but we stress again
that they are respectively detected at about the 95 and 98 per cent
level only. The best--fitting parameters are reported in Table~1 where
we also compute the F--test significance for the most important
spectral components obtained by re--fitting the data after having
excluded the relevant component. In the case of the
disc reflection component, the F--test is computed for the overall
blurred reflection ($99.8\%$), for the addition of the Ni emission
line ($94.8\%$), and for the relativistic blurring only ($99.1\%$)
separately. 

The most important result is that we infer a very strong (and nearly
neutral) reflection component from the accretion disc which dominates the
hard spectrum.  Notice also that the transmitted Fe line width is now
only an upper limit, consistent with an absorber far away from the
nucleus. The transmitted Fe line equivalent width is also slightly
reduced because of the broad Fe line
associated with the disc reflection component. Part of the hard
curvature is now due to the disc reflection component which results
in a slightly smaller column density of the absorber. As for the disc
parameters, the inner disc radius is consistent with both a spinning
and a non--spinning black hole, while the inclination is relatively
poorly constrained.

The large value of the reflection fraction $R=10\pm 3$ (where
$R=\Omega/2\pi$ and $\Omega$ is the solid angle subtended by the
reflector) is unusual and the resulting reflection--dominated spectrum
has consequences on luminosity estimates of the nuclear emission.  If
the spectrum is indeed reflection--dominated, the intrinsic AGN
luminosity is much higher than the observed one. For this reason, in
Table~1 we report two different measures of the 2--10~keV luminosity,
with two different meanings. The first one is the
absorption--corrected observed luminosity ($L^{\rm{obs}}_{2-10}$),
while the second is the estimated intrinsic luminosity of the AGN
($L^{\rm{est}}_{2-10}$): this is the absorption--corrected luminosity
of the direct AGN emission (the hard power law continuum) multiplied by the
reflection fraction i.e.  the observed AGN luminosity plus the
luminosity needed to produce the reflection spectrum we observe. It is
clearly a model--dependent, but in our opinion meaningful, estimate.

\subsection{A partial covering alternative}

In their analysis of the previous {\it BeppoSAX} observation, Dadina
\& Cappi (2004) pointed out that, besides the Compton--thin absorber,
the {\it BeppoSAX} hard spectrum could be described either in terms of
a strong and dominant X--ray reflection component from the disc or in
terms of a partial covering model. Their partial--covering solution
was an attempt to describe the 5--6~keV and $>$10~keV positive
residuals seen in Fig~4 (and visible in the {\it BeppoSAX} data as
well) with the curvature produced by a high--column ($\sim 5\times
10^{24}$~cm$^{-2}$) absorber.  Motivated by their study, we also
considered a similar model and allowed the Fe abundance of the
Compton--thin absorber and of the partial coverer to vary, but forcing
it of course to be the same. We also include the 6.8~keV absorption
line.  However, the partial covering best--fit is statistically worse
than our best--fit model with $\chi^2=470$ for 442 dof (to be
compared with $\chi^2=442$ for 438 dof). The partial--coverer has
$N_H \simeq 5\times 10^{24}$~cm$^{-2}$ and a covering fraction of
about 97 per cent, while the Compton--thin absorber parameters are
consistent with those reported in Table~1. The partial coverer
accounts for the residuals above 10~keV (see Fig.~3) but not for the
positive residuals in the 5--6~keV band and the absorption structure
around 7~keV is only poorly modelled (see Fig.~4). Therefore, we do
not consider the partial covering model any further.

\begin{figure}
\begin{center}
 \includegraphics[width=0.35\textwidth,height=0.455\textwidth,angle=-90]{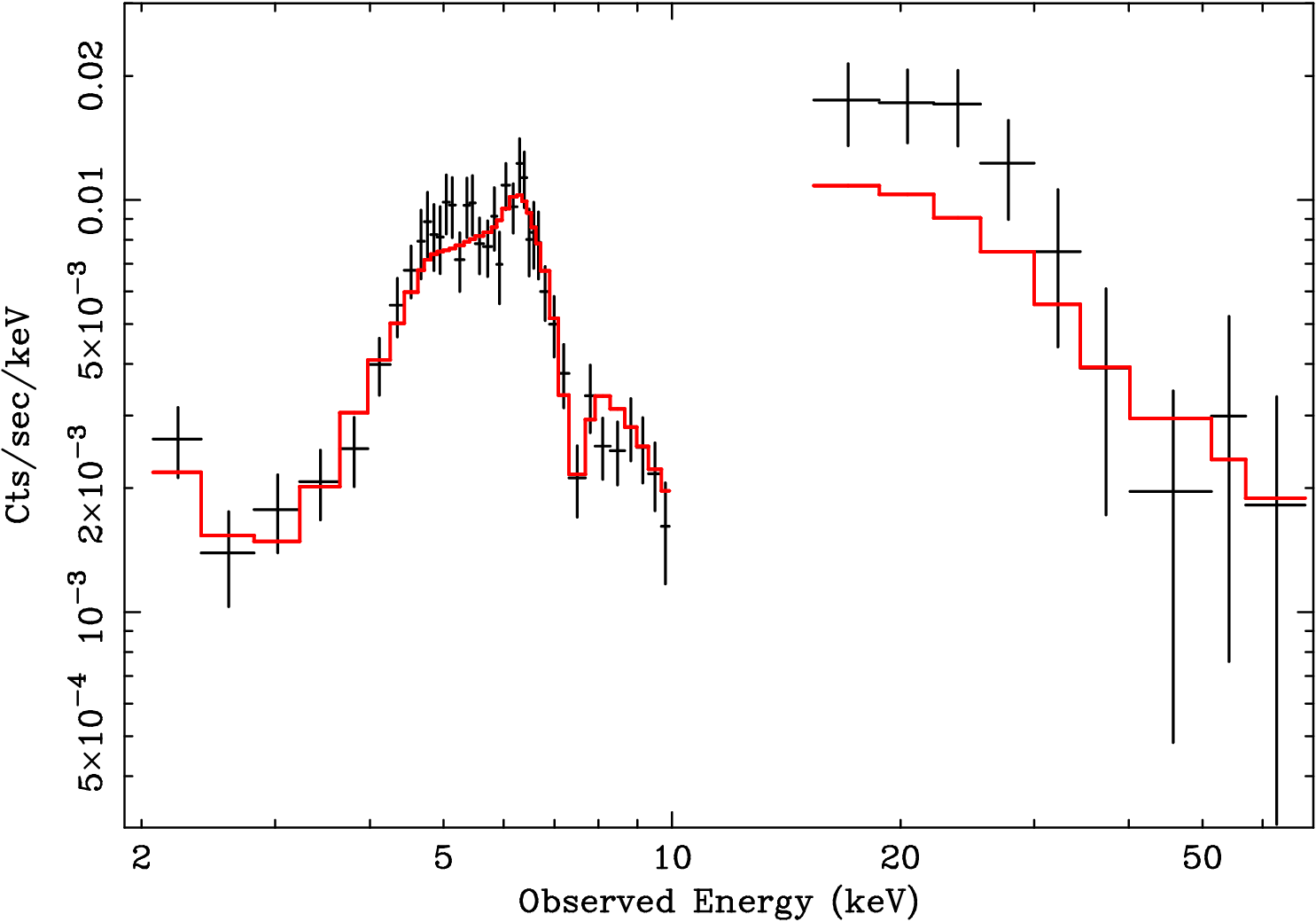}
{\vspace{0.2cm}}
 \includegraphics[width=0.35\textwidth,height=0.455\textwidth,angle=-90]{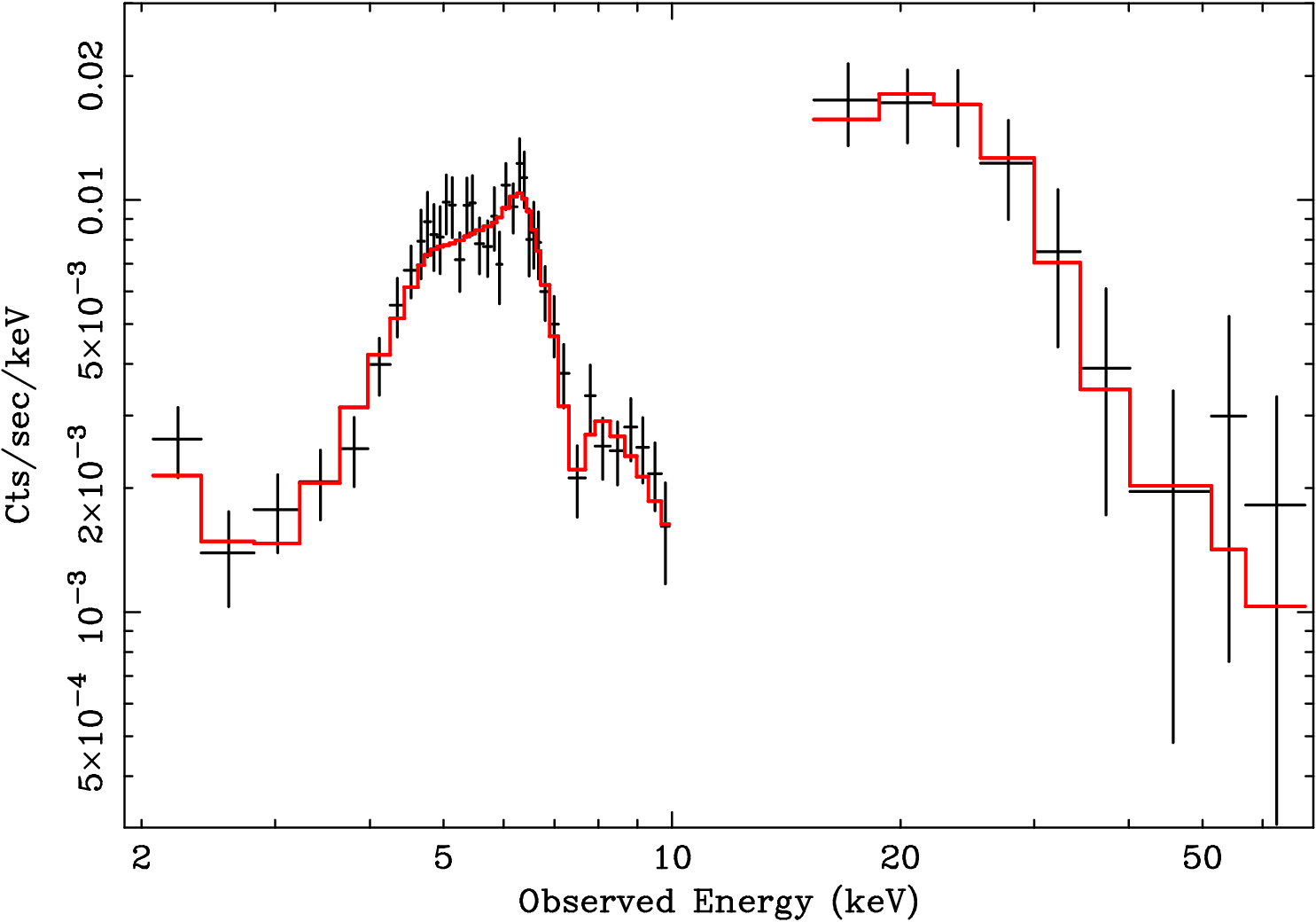}
\end{center}
\caption{In the top panel we show the baseline {\it XMM--Newton} model
  applied to the {\it BeppoSAX} data (MECS and PDS detectors). The fit
  is acceptable but large residuals are seen around 20--30~keV where
  reflection is expected to dominate. In the bottom panel, we add the
  disc reflection component, i.e. we apply the {\it XMM--Newton}
  best--fitting model to the {\it BeppoSAX} data.  }
\end{figure}

\section{The previous {\it ASCA} and {\it BeppoSAX} observations}

IRAS~13197--1627 has been observed twice previously in X--rays, in
July 1995 by {\it ASCA} in the 0.5--10~keV band for 37~ks (Ueno 1997),
and in July 1998 by {\it BeppoSAX} in the 0.5--100~keV band for 44~ks,
reduced to 20~ks above 10~keV (Risaliti 2002; Dadina \& Cappi 2004).
The {\it ASCA} data were retrieved from the Tartarus database and the
event files were processed as standard and verified to give consistent
results with the products obtained by Tartarus team. We used data from
the SIS~0 and 1 and from the GIS~2 and 3 detectors and performed joint
fit to the X--ray spectra in the 2--10~keV band. The {\it BeppoSAX}
data were retrieved from the {\it BeppoSAX} Science Data Centre and
event files were processed as standard. The final MECS and PDS spectra
were found to be in excellent agreement with those from the automated
pipeline (available on--line).

\subsection{Confirming the reflection--dominated spectrum with {\it
    BeppoSAX}}

Besides the broad Fe line and edge, the X--ray reflection model
predicts the presence of a strong Compton hump around 20--30~keV. The
previous {\it BeppoSAX} observation can thus be used to confirm the
detection of the Compton hump in the high--energy PDS detector. We
first apply to the 2--80~keV {\it BeppoSAX} data our baseline model
(no disc reflection). As already reported by Dadina \& Cappi (2004) an
absorption line is tentatively detected at $7.5\pm 0.2$~keV, while
none is seen at 6.8~keV. If related to resonant absorption from
Fe\textsc{xxv} (Fe\textsc{xxvi}), the energy of the line implies an
outflow of at least 27000 (13000)~km~s$^{-1}$. We confirm their result
and add that the line is detected only marginally (at the 98 per cent
level) but if true, its presence could indicate that an outflow with
variable velocity on long timescales is indeed present in
IRAS~13197--1627.

The {\it BeppoSAX} data and baseline (plus absorption line)
best--fitting model are shown in the top panel of Fig.~7. The fit is
acceptable ($\chi^2=76$ for 69 dof), but clear residuals are left
around 20--30~keV, where reflection is expected to dominate. We thus
consider our {\it XMM--Newton} best--fitting model comprising the
X--ray reflection component from the disc. Given the poorer quality of
the data, we fix the Fe abundance, inner disc radius, inclination and
Ni emission line energy to the best--fitting {\it XMM--Newton} values
and fit the {\it BeppoSAX} data with the other parameters free to
vary. The result is shown in the bottom panel of Fig.~7 and shows that
the X--ray reflection component is required by the PDS data. We obtain
a final fit of $\chi^2=52$ for 66 dof, significantly better than the
baseline model one. {\it BeppoSAX} measures a reflection fraction of
$R=11\pm 2$, consistent with the {\it XMM--Newton} one ($R=10\pm
3$). 
\begin{table}
\begin{center}
  \caption{The best--fitting model to the {\it XMM--Newton} data is
    applied to the previous {\it ASCA} (2--10~keV) and {\it BeppoSAX}
    (2--80~keV) observations.  We report the most relevant
    best--fitting parameters and the resulting statistics
    ($\chi^2$/dof). For the {\it BeppoSAX} ({\it XMM--Newton})
    observation, the baseline model also comprises a $\sim$7.5~keV
    ($\sim$6.8~keV) absorption line.}
\begin{tabular}{lccc}          
\hline
{\bf Parameter.} & {\bf{ASCA}} & {\bf{BeppoSAX}} & {\bf XMM--Newton}\\  
& (1995) & (1998) & (2005)\\    
\hline
$\Gamma_h$& $2.6^{+0.5}_{-0.7}$ & $2.2^{+0.7}_{-0.4}$ & $2.5\pm 0.3$\\
$N_H$ & $4.3^{+1.3}_{-0.7}$ & $3.7\pm 0.8$& $3.9\pm 0.4$\\
$E_{Fe}$ & $6.5\pm 0.1$ & $6.4\pm 0.1$& $6.40\pm 0.02$\\
$N_{Fe}$ & $6.5\pm 0.9$ & $6.8\pm 0.8$& $6.2\pm 0.4$\\
\hline
$F_{\rm{cont}}$ & $9.2 \pm 1.6$ & $11.4 \pm 1.3$& $6.8^{+1.9}_{-2.3}$\\
$F_{\rm{refl}}$ & $7.0\pm 2.7$ & $8.0\pm 1.1$& $3.9\pm 0.8$\\
\hline
Best Fit & 156/150& 52/66 & 442/438\\
Baseline & 165/153& 76/69 & 505/445\\
\hline
\end{tabular}
\end{center}
The column density of the absorber ($N_H$) is in units of
$10^{23}$~cm$^{-2}$ and the Fe line energy ($E_{Fe}$) is in
keV. The Fe line normalisation ($N_{Fe}$) is in units of
$10^{-6}$~ph~cm$^{-2}$~s$^{-1}$. The continuum and reflection fluxes
($F_{\rm cont}$ and $F_{\rm refl}$)
are unabsorbed and given in the
2--10~keV band in units of $10^{-12}$~erg~cm$^{-2}$~s$^{-1}$. 
\end{table}

\subsection{Long--term spectral variability}

We have studied the long--term spectral variability of
IRAS~13197--1627 by using the three available X--ray observations with
the goal of determining i) changes in the continuum slope, column
density of the absorber, and narrow Fe line flux, and ii) variability
of the two main spectral components, namely the continuum power law
and the disc reflection. We mention here that, based on literature
results from {\it ASCA} (Ueno 1997) and on the Risaliti (2002)
analysis of the {\it BeppoSAX} data, Risaliti, Elvis \& Nicastro
(2002) have pointed out that a long--term variability in the absorbing
column density could be claimed in IRAS~13197--1627, which would be
consistent with a clumpy absorbing medium close to the central engine.

We applied our best--fit model to the {\it ASCA} spectrum as well and
our results are reported in Table~2 for the three observations.  The
source observed (i.e. absorbed) 2--10~keV flux was low during the {\it
  XMM--Newton} observation ($2.1\pm 0.2\times
10^{-12}$~erg~cm$^{-2}$~s$^{-1}$), about a factor 2 higher during the
{\it BeppoSAX} one ($4.7 \pm 0.7\times
10^{-12}$~erg~cm$^{-2}$~s$^{-1}$), and intermediate during the {\it
  ASCA} pointing ($3.5 \pm 0.9\times
10^{-12}$~erg~cm$^{-2}$~s$^{-1}$). Table~2  shows that the hard
power law photon index, the column density of the Compton--thin
absorber, and the transmitted Fe line normalisation are all consistent
with being the same in the three observations. We conclude that the
data collected so far are consistent with the flux variability
occurring at constant spectral slope and absorption, meaning that we
have no observational evidence to support the conclusion that the
absorber is located relatively close to the central engine (although
we cannot exclude it).
\begin{figure}
\begin{center}
 \includegraphics[width=0.35\textwidth,height=0.455\textwidth,angle=-90]{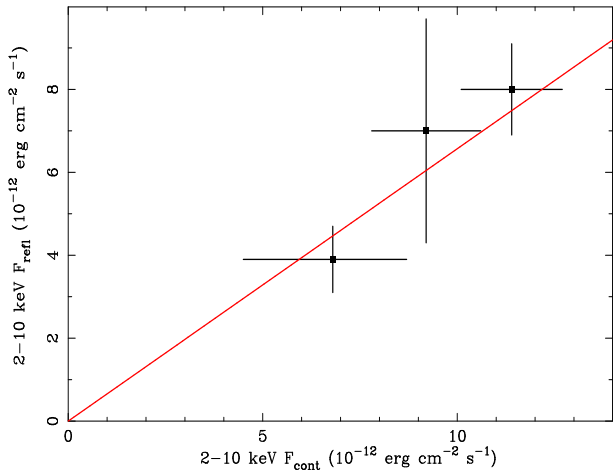}
\end{center}
\caption{The 2--10~keV reflection flux is plotted against the
  2--10~keV continuum flux from the three available observations. This
  demonstrates that the reflection component responds to the continuum
  long--term variability within the errors. We also show as a reference
  the best--fitting linear relationship.
}
\end{figure}

As for the variability of the continuum and reflection components, we
find that they are consistent with being correlated. This is shown in
Fig.~8 where the 2--10~keV flux of the reflection component is plotted
against the 2--10~keV flux of the continuum (both unabsorbed). The
origin of the reflection--dominated spectrum and of its variability
are discussed below

\section{Origin of the reflection--dominated spectrum}

The most remarkable result of our analysis is the presence of a
dominant reflection component from the accretion disc with a
reflection fraction of $R=10\pm 3$. In other words, the hard X--ray
spectrum of IRAS~13197--1627 is largely reflection--dominated. Since
reflection comes from the inner accretion disc, and since most of the
accretion power has to be radiated from there (or in a corona above
it), it is very difficult to imagine a situation in which the spectrum
is reflection--dominated because absorption covers the direct
continuum X--ray source and not the inner disc. We tried anyway a
spectral model in which the hard power law and the reflection
component are absorbed by different column densities to check whether
this could help explaining the large reflection fraction. However, we
did not find any statistical improvement and, more importantly, the
two column densities turned out to be consistent with each other and
with the value reported in Table~1. 

One possibility is that the large observed reflection fraction is due
to strong gravity effects. If the primary emitting source of X--rays
is located only a few $r_g$ from the black hole, light bending focus
the continuum towards the accretion disc (and black hole) dramatically
reducing the primary continuum at infinity and enhancing the
reflection fraction up to the very large observed values (see the
light bending model proposed by Miniutti et al.  2003; Miniutti \&
Fabian 2004; see also Fabian et al 2005; Ponti et al 2006; Miniutti et
al 2006 for recent applications).  As for the variability, the model
was devised to reproduce large continuum variation with no or little
reflection variability (as observed e.g. in MCG--6-30-15, see Fabian
\& Vaughan 2003; Miniutti et al 2006). The model successfully
reproduces this behaviour for sources with $R$ between $\sim 1$ and
$\sim 3-4$ (such as MCG--6-30-15), while it predicts a correlated
variability for reflection--dominated sources with $R>3-4$. Thus, in the
present case, the continuum/reflection correlated variability (see
Fig.~8) is consistent with the model (since $R\sim 10$).  However, a
correlation between disc reflection and continuum is expected in any
simple disc reflection model and is not specific of the light bending
one. The main advantage of the model is that the large reflection
fraction is naturally explained.

As recently pointed out by Merloni et al. (2006),
reflection--dominated spectra could also be produced from discs which
are subject to instabilities making both density and heating rate in
the inner disc inhomogeneous. If the clouds generated by the
inhomogeneous flow have high effective optical depth, steep spectra
($\Gamma \sim 2.4$) emerge and they are associated with a dominant
reflection component, in good agreement with our analysis. It is a
natural consequence of this model that some of the inner--disc clouds
responsible for the X--ray reflection would be seen in absorption with
a range of column densities (say $10^{23}-10^{25}$~cm$^{-2}$).  Given
the location of the clouds (few innermost $r_g$) short--timescale
variability in absorption has to be expected and could be used to
disentangle between the two models in future longer observations. At
the present time, we have no evidence for absorption variability
(notice that the Compton--thin absorber we detect has nothing to do
with this scenario since the transmitted Fe line is too narrow to be
produced in the inner--disc).

\section{The rich soft X--ray spectrum}

Since the source primary emission is relatively heavily obscured by a
$\sim 4 \times 10^{23}$~cm$^{-2}$ column density, the AGN emission
does not contribute directly in the soft band. This is typical of
Seyfert~2 galaxies in which the soft emission is mainly due to a blend
of emission lines for He--like and H--like K transitions of light
elements and L transitions of Fe (e.g. Kinkhabwala et al. 2002; Bianchi
et al. 2005a). The emission lines are most likely from gas
photo--ionized by the AGN and located on large scales (possibly
coincident with the Narrow--Line region, see e.g. Bianchi, Guainazzi
\& Chiaberge 2006) and/or collisionally ionized plasmas (e.g.
associated with star--forming regions or gas shocked by a jet).  In
both cases a residual weak continuum due to electron--scattering of
the primary AGN radiation by extended gas  or to part of the
nuclear emission leaking through the absorber can be present. We first
concentrate on the high--resolution RGS spectrum and search for
signatures of emission lines related to photo--ionized gas and/or
thermal plasmas emission.

\subsection{The high--resolution RGS data}

\begin{figure}
\begin{center}
{
\includegraphics[width=0.35\textwidth,height=0.455\textwidth,angle=-90]{Olines.cps}
{\vspace{0.2cm}}
\includegraphics[width=0.35\textwidth,height=0.455\textwidth,angle=-90]{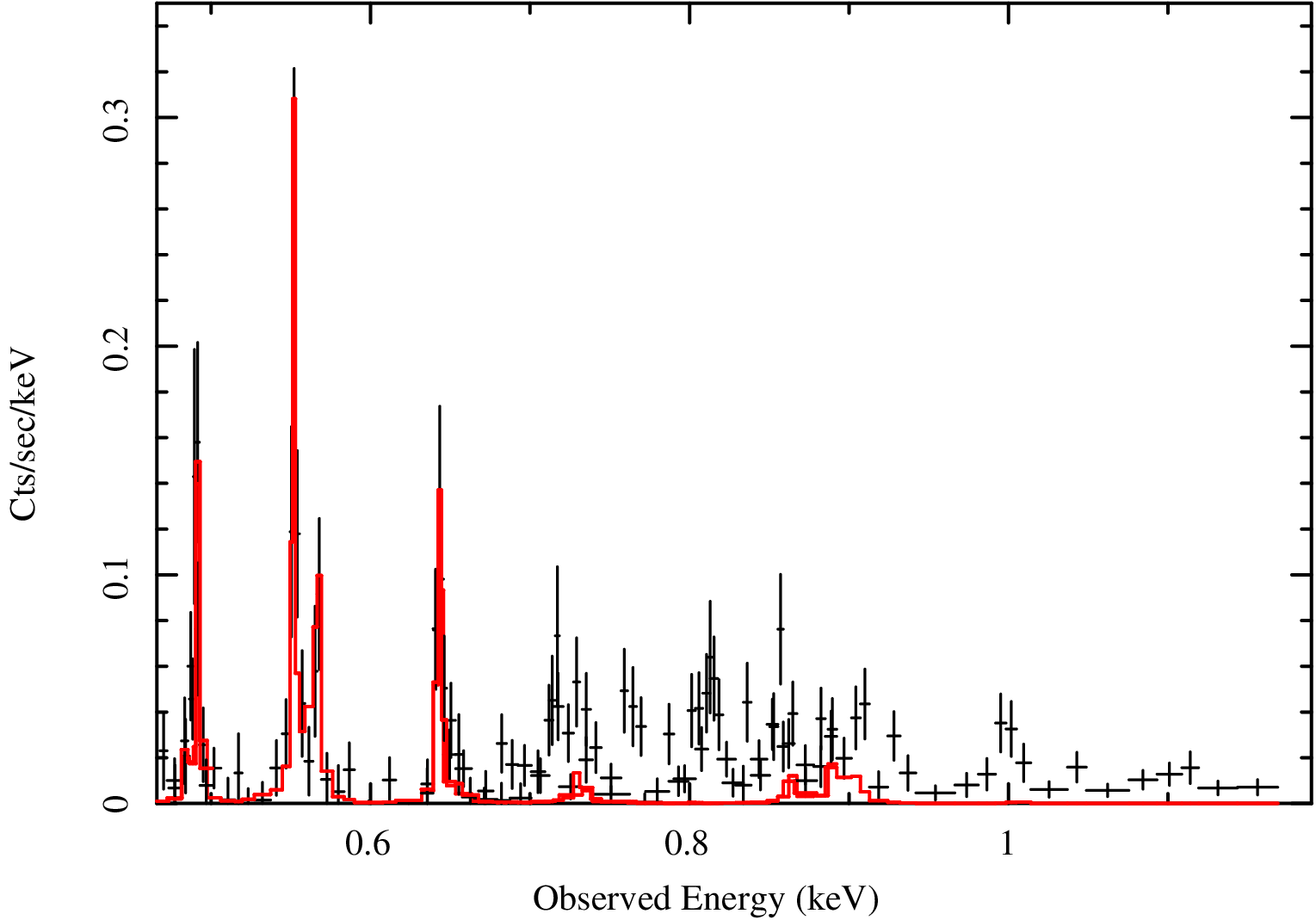}
}
\end{center}
\caption{{\bf Top:} a portion of the first order RGS spectrum in the
  0.45--0.7~keV band. The vertical lines are the theoretical energies
  of the emission lines shifted to account for the source redshift,
  while the solid line is the best--fitting model comprising a power
  law absorbed by the Galactic column plus four Gaussian emission
  lines.  The dominant line in the He--like O~VII multiplet is the
  forbidden, suggesting emission from photo--ionized plasma as
  typically observed in Seyfert 2 galaxies. {\bf Bottom:} the RGS data
  up to 1.2~keV are shown together with the best--fitting {\tt XSTAR}
  model. Photo--ionized gas emission alone does not seem to be able to
  reproduce the emission lines seen above $\sim$0.7~keV.}
\end{figure}

\subsubsection{Photo--Ionized plasma} 

The first order 0.45--0.7~keV RGS spectrum is shown in the top panel
of Fig.~9 where clear K$\alpha$ emission lines from N~{\textsc{vii}},
O~{\textsc{vii}}, and O~{\textsc{viii}} are seen (the O~{\textsc{vii}}
line is actually a He--like multiplet comprising forbidden,
intercombination, and resonance lines).  In all subsequent fits, a
phenomenological power law continuum absorbed by the Galactic column
density is considered and its properties will be discussed in detail
in the subsequent sections.

We detect K$\alpha$ emission lines from N~{\textsc{vii}} at $0.50\pm
0.05$~keV and O~{\textsc{viii}} at $0.653\pm 0.006$~keV, while the
O~{\textsc{viii}} k$\beta$ line is only an upper limit. As for the
O~{\textsc{vii}} He--like multiplet, we clearly detect a strong
forbidden (f) line at $0.561\pm 0.002$~keV and a fainter resonance (r)
one at $0.573\pm 0.004$~keV. All observed energies are consistent with
the rest--frame theoretical energies and are reported together with
the corresponding fluxes in Table~2. The O~{\textsc{vii}}
forbidden/resonance flux ratio is $2.4^{+1.9}_{-0.9}$. The dominance
of the O~{\textsc{vii}} forbidden line relative to the resonance is
generally taken as a signature of emission predominantly from
photo--ionized rather than collisional plasmas (e.g.  Porquet \& Dubau
2000 and references therein) and we conclude that such a component is
indeed present in the soft X--ray spectrum of IRAS~13197--1627. We
also modelled the data with the photo--ionized emission code {\tt
  XSTAR} and find a good fit ($\chi^2=53$ for 46 dof) with a column
density of $\sim 1\times 10^{22}$~cm$^{-2}$ and ionization $\log\xi
\sim 1$~erg~cm~s$^{-1}$ (both loosely constrained).  The power law
continuum is (although marginally) required by the data at the 96 per
cent level (F--test), but its slope and normalisation are largely
unconstrained because most of the RGS photons are in the emission
lines rather than in the continuum. The issue will be re--examined
when the soft EPIC spectrum is discussed below.

\subsubsection{Collisionally--ionized plasma emission?} 

When higher energy data up to 1.2~keV are considered, large residuals
are seen and several additional emission lines are required above
0.7~keV (see Fig.~9, bottom). Re--fitting the {\tt XSTAR} model
discussed above in the 0.45--1.2~keV band produces an unacceptable fit
with $\chi^2=263$ for 127 dof. We then add further Gaussian emission
lines as required in order to account for the residuals seen above
0.7~keV. The lines we detect are reported in Table~2 together with
their most likely identification and with the properties of the power
law continuum (see caption). We clearly detect Ne~{\textsc{ix}} and
Ne~{\textsc{x}} K$\alpha$ lines and a plethora of other emission
features from Fe L. The Fe L emission complex (mostly from
Fe~{\textsc{xvi}} and Fe~{\textsc{xvii}}) is generally interpreted as
a sign of collisionally--ionized plasma emission. However, as pointed
out e.g.  by Kinkhabwala et al. 2002 (and many others) ambiguities
between photo--ionized gas and thermal plasma emission are difficult
to disentangle. One possibility is represented by the detection of
radiative recombination continua (RRC) or higher order series
transitions for O~{\textsc{vii}} ($\beta, \gamma, \delta$) which would
rule out a very strong contribution from a thermal plasma (e.g.
Kinkhabwala et al. 2002). In the present case, however, the statistics
in the RGS is limited and we cannot easily disentangle between the two
possibilities. In the following we present a possible solution in
terms of photo--ionized plus collisionally--ionized emission, but firm
conclusions must await better quality high--resolution spectra.

\begin{table}
  \caption{\label{RGSlines}List of the emission lines included in the
    best fit model for the RGS data between 0.45~keV and 1.2~keV. The 90
    per cent errors are given in parenthesis on the last digit. When
    required, the upper level (UL) and lower level (LL) of the most
    likely transition are reported. The model also comprises a poorly
    constrained continuum, modelled with a power law with 
    $\Gamma_s = 2.1\pm 0.8$ and normalisation of $3.5\pm 3.0 \times
    10^{-5}$~ph~cm$^{-2}$~s$^{-1}$ (we assume Galactic absorption).}
\begin{center}
\begin{tabular}{cccc}
E$_{\rm obs}$(keV) & Flux$^a$ & Likely Id. &
E$_{\rm{theo}}$ (keV)\\
& & UL& LL\\
\hline
& & & \\
$0.50(5)$ & $1.8(6)$ & {N\,\textsc{vii}} K$\alpha$ & 0.500\\
& & & \\
$0.561(2)$ & $3.8(9)$ & {O\,\textsc{vii}} K$\alpha$ &0.561 (f)\\
& & & \\
$0.573(4)$ & $1.6(5)$ &{O\,\textsc{vii}} K$\alpha$ & 0.569 (i); 0.574 (r)\\
& & & \\
$0.653(6)$ & $2.0(5)$ & {O\,\textsc{viii}} K$\alpha$& 0.654\\
& & & \\
$0.666^*$ & $< 1$ &{O\,\textsc{vii}} K$\beta$ & 0.666\\
& & & \\
$0.73(2)$ & $1.0(5)$ & {Fe\,\textsc{xvi}} 
 & 0.727\\
&  & $2p^2_{1/2} 2p^3_{3/2} 3s$&$2p^2_{1/2} 2p^4_{3/2}$\\
&&&\\
$0.74(2)$ & $1.0(7)$ & {Fe\,\textsc{xvi}}  &  0.739\\
&   &  $2p^2_{1/2} 2p^4_{3/2} 3s$&$2p^2_{1/2} 2p^4_{3/2}$\\
& & & \\
$0.78(2)$ & $0.8(4)$ & {Fe\,\textsc{xvii}} &  0.775\\
& &  $2p^2_{1/2} 2p^2_{3/2} 3s$&$2p^2_{1/2} 2p^3_{3/2}$\\
& & & \\
$0.827(3)$ & $1.3(5)$ & {Fe\,\textsc{xvi}}
& 0.826\\
& & $2p_{1/2} 2p^4_{3/2} 3d_{3/2}$&$2p^2_{1/2} 2p^4_{3/2}$\\
& & & \\
$0.85(2)$ & $1.0(4)$ & {Fe\,\textsc{xvii}} &  0.851\\
& &  $2p^2_{1/2} 2p^2_{3/2} 3d_{5/2}$&$2p^2_{1/2} 2p^3_{3/2}$\\
& & & \\
$0.870(3)$ & $1.1(5)$ & {Fe\,\textsc{xvii}} & 0.867
\\
& & $2p_{1/2} 2p^3_{3/2} 3d_{3/2}$&$2p^2_{1/2} 2p^3_{3/2}$\\
& & & \\
$0.90(1)$ & $0.7(3)$ & {Ne\,\textsc{ix}} K$\alpha$ &  0.905 (f)\\
& & & \\
$0.92(1)$ & $1.3(5)$ & {Ne\,\textsc{ix}} K$\alpha$ &
0.915 (i); 0.922 (r)\\
& & & \\
$1.02(2)$ & $0.9(6)$ & {Ne\,\textsc{x}} K$\alpha$&  1.022\\
& & & \\
$1.11(2)$ & $0.8(5)$ & {Fe\,\textsc{xvi}} & 1.112\\
& & $2p_{1/2} 2p^4_{3/2} 5d_{3/2}$&$2p^2_{1/2} 2p^4_{3/2}$\\
& & & \\
\hline
\end{tabular}
\end{center}
$^a$ Fluxes in units of $10^{-5}$ ph cm$^{-2}$ s$^{-1}$.

$^*$ Indicates that the parameter has been fixed.
\end{table}

\subsubsection{A global model to the RGS data}

A good quality global fit to the 0.45--1.2~keV RGS data can be
obtained by considering a composite model comprising the X--ray
continuum (power law absorbed by the Galactic column), a photo--ionized
gas emission component as obtained with the {\tt XSTAR} code, and a
thermal plasma component modelled through the {\tt VMEKAL} model to
account for the Fe L complex. Given that the 2--12~keV spectrum
provides indication that Fe is overabundant, we let the Fe abundance
in the {\tt VMEKAL} and {\tt XSTAR} models free to vary and tied to
each other, while all the other elements have fixed solar abundances.
We obtain a good global fit ($\chi^2=153$ for 125 degrees of freedom).
The higher energy lines (see Fig.~9) are well accounted for by the
thermal model and we measure a temperature of
$0.63^{+0.04}_{-0.05}$~keV, while the Fe abundance is
$1.3\pm 0.3$ times solar. The {\tt VMEKAL} model improves the
fit by $\Delta\chi^2=110$ for 3 dof with respect to the power law plus
{\tt XSTAR} spectral model. We point out that we were not able to
obtain a similar improvement by considering two photo--ionized
components with different ionization parameters and column densities
suggesting that a non photo--ionized component is indeed required by
the data. Some positive residuals are still left around 1~keV and they
can be modelled by adding a second higher--temperature thermal
component.  Its temperature is only poorly constrained in the range of
1.5--4~keV but the additional model improves further the statistics
with $\Delta\chi^2=20$ for 2 more free parameters.  According to our
best--fitting model extrapolated to 2~keV, the photo--ionized component
has a 0.5--2~keV luminosity of $5\pm 2 \times 10^{40}$~erg~s$^{-1}$,
while the two thermal components dominate the soft band with a
luminosity of $1.4\pm 0.4 \times 10^{41}$~erg~s$^{-1}$.  The soft
power law contribution is much more uncertain ($8\pm 7 \times
10^{40}$~erg~s$^{-1}$) and will be better constrained with the higher
statistics EPIC spectrum as discussed below.

\begin{figure}
\begin{center}
{
 \includegraphics[width=0.35\textwidth,height=0.455\textwidth,angle=-90]{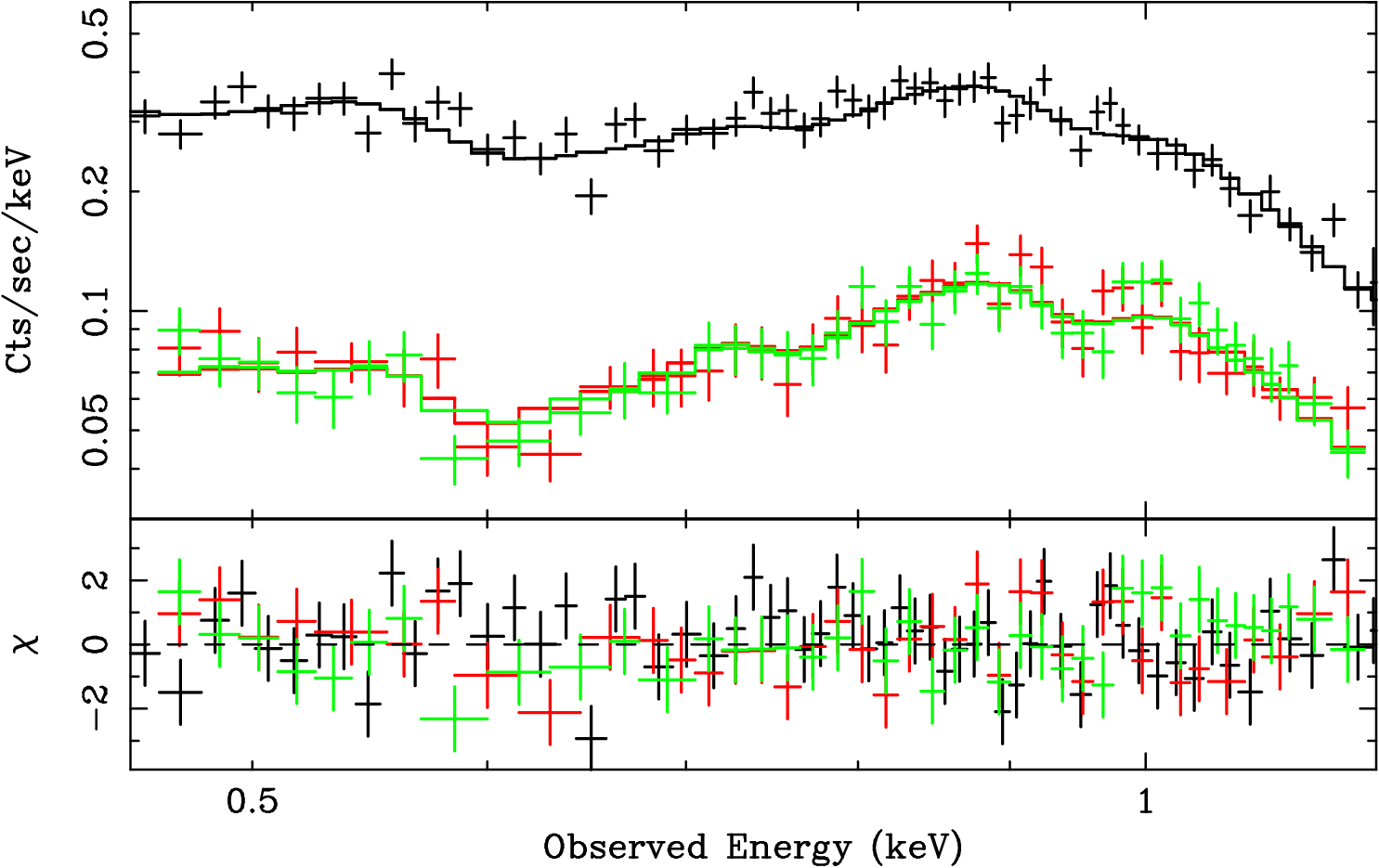}
{\vspace{0.2cm}}
 \includegraphics[width=0.35\textwidth,height=0.455\textwidth,angle=-90]{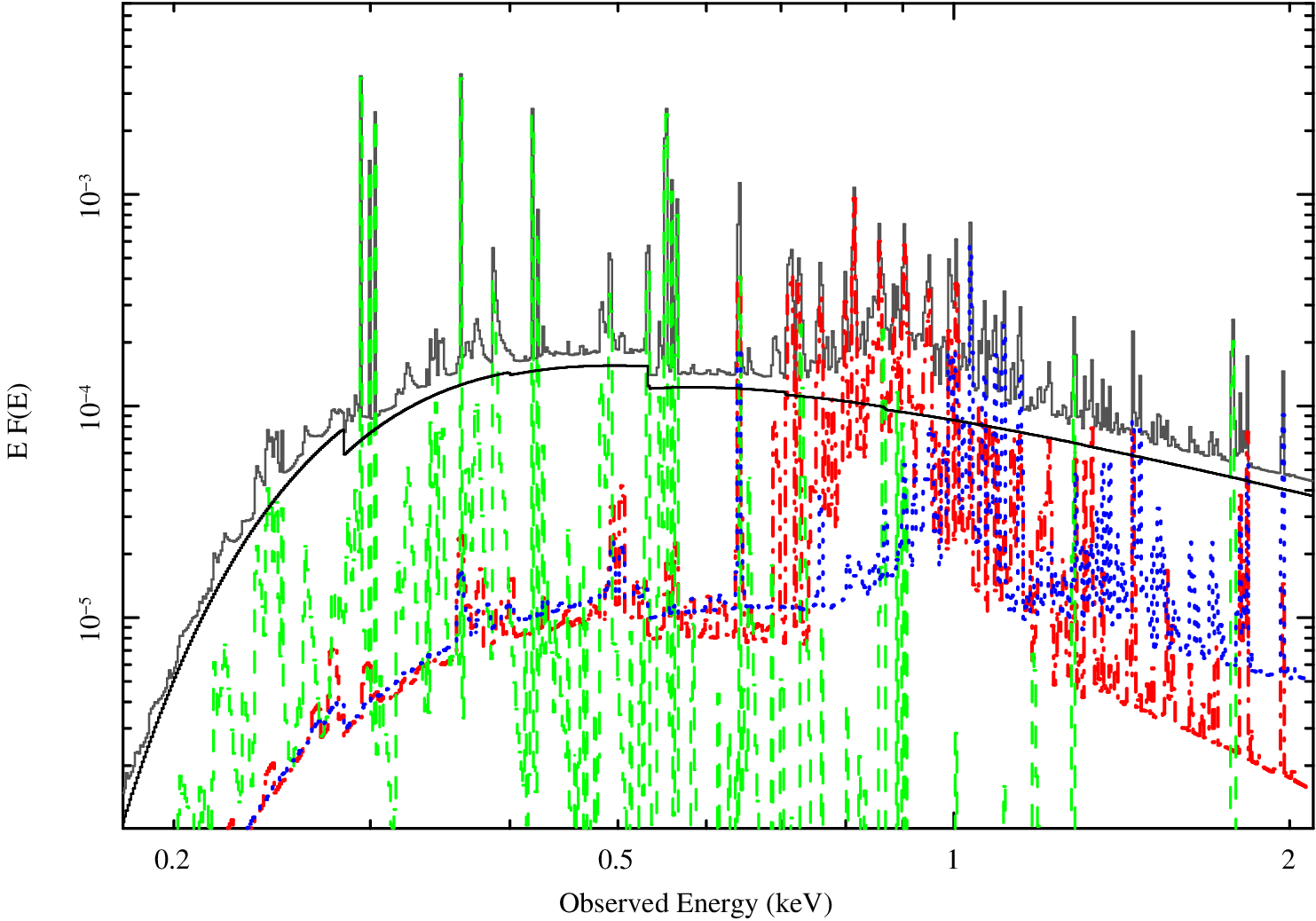}
}
\end{center}
\caption{{\bf Top:} the 0.45--1.2~keV EPIC pn and MOS spectra fitted
  with the RGS model comprising a power law continuum,
  emission from photo--ionized gas modelled through the {\tt XSTAR} code
  and two thermal components ({\tt VMEKAL}).  {\bf Bottom:} the
  best--fitting model is shown in the 0.2--2~keV band (grey).  Besides
  the soft power law, the photo--ionized gas  and the two
  thermal plasma spectral models  are shown. The
  overall spectrum is absorbed by the Galactic column density.}
\end{figure}

\subsection{The CCD--resolution EPIC data}

We have applied our best--fit RGS model to the pn and MOS data in the
same 0.45--1.2~keV energy band. This allows us to i) check the
consistency between the high--resolution and CCD--resolution spectra,
and ii) explore whether the soft power law continuum is truly required
by the data. We obtain a good description of the EPIC data with a
final result of $\chi^2=273$ for 233 dof with a two--temperature model
with $0.6\pm 0.2$~keV and $1.4\pm 0.2$~keV plus photo--ionized gas
emission. The Fe abundance is measured from the {\tt XSTAR} and {\tt
  VMEKAL} models as $1.4^{+0.5}_{-0.3}$ times solar which is fully
consistent with the Fe overabundance we found modelling the 2--12~keV
hard spectrum (see Table~1). However, if the Fe abundance is forced to
be solar, the worsening of the statistics is marginal and a
super--solar abundance is required at the 95 per cent level only.
Finally, the power law continuum is very significantly required by the
data: if it is excluded from the model, the statistics is worse by
$\Delta\chi^2= 32$ (2 dof). We measure a slope $\Gamma_s =
2.3\pm 0.3$ and a normalisation of $7.0 \pm 2.0 \times
10^{-5}$~ph~cm$^{-2}$~s$^{-1}$. Our results are summarised in Table~4
and remain unchanged in the broadband analysis discussed below. The
EPIC spectra, best--fit model and data to model ratio are shown in the
top panel of Fig.~10, while the best--fitting model components are
shown in the bottom panel.

\begin{table}
  \caption{Parameters of the best--fit model to the 0.45--1.2~keV EPIC
    spectrum. The model comprises Galactic absorption, a power law
    continuum, two thermal
    components ({\tt VMEKAL}) and a photo--ionized gas emission one
    ({\tt XSTAR}). The Fe abundance is forced to be the same in the two
    thermal and the photo--ionization models. For each component we also
    report the luminosity in the 0.5--2~keV band (we checked these
    luminosities also in the broadband analysis discussed below and
    find excellent agreement). All components are statistically
    very significantly required. However, forcing a solar Fe abundance
    produce a worsening of the fit at the $\sim 95 \%$ level only.}
\begin{center}
\begin{tabular}{lcr}
  \hline
  {\bf{Parameter}} & {\bf{Value}} & {\bf F--test}\\
  \hline
  \multicolumn{2}{l}{\bf{Continuum}} &  $>99.99\%$\\
  \hline
  $\Gamma_s$& $2.3\pm 0.3$ & \\
  $L^{\rm{s}}$~[$10^{41}$~erg~s$^{-1}$] & $1.0\pm 0.3$ &\\
  \hline
  \multicolumn{2}{l}{\bf{VMEKAL~1}} &  $>99.99\%$\\
  \hline
  $KT_1$~[keV] & $0.6\pm 0.2$ &\\
  $L^{\rm{th}}_1$~[$10^{41}$~erg~s$^{-1}$] & $0.8\pm 0.2$ &\\
  \hline
  \multicolumn{2}{l}{\bf{VMEKAL~2}} &  $>99.99\%$\\
  \hline
  $KT_2$~[keV] & $1.4\pm 0.2$ &\\
  $L^{\rm{th}}_2$~[$10^{41}$~erg~s$^{-1}$] & $0.6\pm 0.2$ &\\
  \hline
  \multicolumn{2}{l}{\bf{XSTAR}} &  $>99.99\%$\\
  \hline
  $N_H$~[$10^{22}$~cm$^{-2}$] & $0.9\pm 0.5$ &\\
  $\log\xi$~[erg~cm~s$^{-1}$] & $0.8\pm 0.4$ &\\
  $L^{\rm{ph}}$~[$10^{41}$~erg~s$^{-1}$] & $0.4\pm 0.2$ &\\
  \hline
  \multicolumn{2}{l}{\bf{Fe abundance}} &  $>94.8\%$\\
  \hline
  $A_{\rm Fe}$~[solar] & $1.4^{+0.5}_{-0.3}$  &\\
  \hline
\end{tabular}
\end{center}
\end{table}

\section{The broadband 0.45--12~keV spectrum}

After having analysed the X--ray spectrum of IRAS~13197--1627 by
looking separately at the different energy bands and available energy
resolutions, the task of describing the broadband  EPIC
spectrum of the source is greatly simplified. We consider here the
best--fit reflection--dominated model to the hard spectrum (Table~1)
and add to it the soft model we just discussed above. The Fe abundance
in the Compton--thin absorber, the reflection component, the two
thermal models, and the photo--ionized gas one is forced to be the
same for consistency. We find an excellent description of the
broadband 0.45--12~keV EPIC data with $\chi^2=924$ for 820 dof. The spectra,
best--fit model, and residuals are shown in Fig.~11. The model
components are indistinguishable from those shown in the bottom panels
of Fig~6 (hard band) and Fig.~10 (soft band). 

The best--fitting parameters for the reflection component,
relativistic blurring, 6.8~keV absorption line, photo--ionized and
collisional plasma emission are consistent with Table~1 (hard band)
and with the soft X--rays modelling (Table~4). The Fe
abundance is only slightly super--solar ($1.4\pm 0.3$ times solar). By
including the hard band however, a super--solar Fe abundance is
required at the 97 per cent level (see results in Table~1 and 4 for
the hard and soft band respectively). The soft and hard power law
slopes are consistent with each other with $\Gamma_s=2.3\pm 0.3$ and
$\Gamma_h=2.5\pm 0.3$, and the soft power law is required at more than
the 99.99 per cent level.  The hard power law has a normalisation of
$5.7^{+1.6}_{-1.8}\times 10^{-3}$~ph~cm$^{-2}$~s$^{-1}$, while the
soft one has $7.0 \pm 2.0\times 10^{-5}$~ph~cm$^{-2}$~s$^{-1}$.
It should be stressed that the soft power law contributes by only
$\sim$36 per cent to the 0.5--2~keV flux, the remaining $\sim$64 per
cent being due to photo--ionized gas ($\sim$14 per cent) and thermal
plasma emission ($\sim$50 per cent).  The possible origins for the
soft power law--like continuum component are i) scattered nuclear
continuum e.g.  from the same photo--ionized/collisionally--ionized
gas seen in emission in the RGS data, or ii) transmitted nuclear continuum (plus
reflection) through a partial covering absorber (see e.g. Levenson et
al. 2006 and references therein for a discussion). In both cases, we
expect less fractional variability in the soft than in the hard band
because i) the scattering medium is most likely
extended or ii) the transmitted continuum represents 36 per cent only
of the soft flux which is dominated by extended emission. In both
cases the variability would be washed out in the soft band much more
than in the hard, as observed (see Fig~1 and relative discussion).
Strictly speaking the first scenario would not allow for any short
timescale variability in the soft X--ray band (and none is detected,
see Fig.~1).

\begin{figure}
\begin{center}
\includegraphics[width=0.35\textwidth,height=0.455\textwidth,angle=-90]{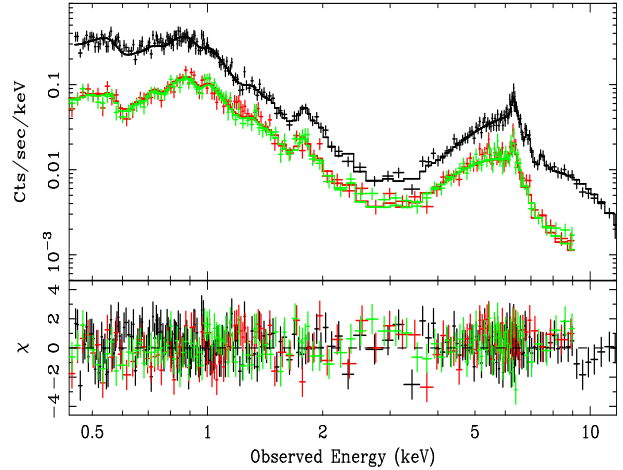}
\end{center}
\caption{The broadband EPIC spectra are shown together with the
  best--fitting spectral model. The model components are as shown in
  Fig.~6 (bottom) and Fig.~10 (bottom) for the hard and soft band
  respectively.
}
\end{figure}

To explore the second solution, we allow the Compton--thin absorber to
be a partial coverer. The soft power law is removed from the spectral
model in the attempt of reproducing it with the fraction of nuclear
emission (hard power law plus reflection) leaking through the
absorber.  We find a solution with the same statistical quality as
with the soft power law model ($\chi^2=922$ for 820 dof). The absorber
has $N_H = 4.1\pm 0.4\times 10^{23}$~cm$^{-2}$ and covering fraction
$C_f =98.5 \pm 1.1$ per cent.  However, we also need a further column
of $9.5\pm 2.5 \times 10^{20}$~cm$^{-2}$ (in excess of the Galactic
value) covering totally the nucleus. We conclude that the soft X--ray
band is dominated by emission lines from both photo--ionized gas and
thermal plasmas and that the nature of the remaining $\sim$36 per cent
continuum cannot be constrained: a power law continuum is
statistically equivalent to a $\sim$1--2 per cent AGN emission (direct
continuum plus reflection) leaking through the absorber.

The observed 0.5--2~keV (2--10~keV) flux is $3.6\pm 1.2\times
10^{-13}$~erg~cm$^{-2}$~s$^{-1}$ ($2.1\pm 0.2\times
10^{-12}$~erg~cm$^{-2}$~s$^{-1}$). Our best--fitting global model also
allows us to compute the absorption--corrected luminosity of the
different spectral components (see also Table~1 and 4). The 2--10~keV
observed luminosity is $L^{\rm{obs}}_{2-10} = 7\pm 2 \times
10^{42}$~erg~s$^{-1}$ only but, if our interpretation of the hard
X--ray spectrum is correct, the estimated AGN luminosity in the
2--10~keV band turns out to be $L^{\rm{est}} = 4.2^{+2.8}_{-2.2}
\times 10^{43}$~erg~s$^{-1}$ (see Table~1 and text). As reported in
Table~4, the total 0.5--2~keV luminosity of the two thermal components
is $L^{\rm{th}} = 1.4 \pm 0.4 \times 10^{41}$~erg~s$^{-1}$, while the
photo--ionized gas contributes with $L^{\rm{ph}} = 4 \pm 2 \times
10^{40}$~erg~s$^{-1}$ in the same band.  Finally, the soft X--ray
continuum (scattered or leaking through the absorber) has a luminosity
of $1.0\pm 0.3 \times 10^{41}$~erg~s$^{-1}$ in the 0.5--2~keV band.

\section{Discussion}

IRAS~13197--1627 hosts a Compton--thin AGN in which the primary X--ray
nuclear emission is steep ($\Gamma_h=2.5\pm 0.3$) and is absorbed by a
column density of $\sim 4\times 10^{23}$~cm$^{-2}$ of neutral gas. A
neutral Fe K$\alpha$ emission line is also clearly detected at 6.4~keV
and is consistent with being transmitted through the absorber which
probably has a slightly super--solar Fe abundance. The column density
inferred from the X--rays is at odds with previous optical studies.
From the work by Aguero et al (1994) we infer E(B-V)$\simeq 0.47$,
while Cid Fernandes et al (2004) measure A$_V \simeq 0.38$ (note that
the two values do not satisfy the standard conversion factor
A$_V$/E(B-V)$=3.1$). As already pointed out by Maiolino et al (2001a;
2001b) E(B-V)/$N_H \simeq 1.2\times 10^{-24}$, about two orders of
magnitude smaller than the typical Galactic value ($1.7\times
10^{-22}$). This huge discrepancy is not uncommon in AGNs (Maiolino et
al 2001a) and could be explained by the presence of large dust grains
in the circum--nuclear region of AGNs (e.g.  due to grain coagulation in
dense clouds), which results in a featureless extinction curve with
reduced E(B-V)/$N_H$. The large grain/coagulation scenario would also
significantly reduce the A$_V/N_H$ ratio, as observed here. Other
possibilities include dust sublimation if the X--ray absorber is
sufficiently close to the luminous nucleus, the idea that most of the
X--ray absorption is due to the same clouds responsible for the broad
line region emission, and/or a very low dust--to--gas mass ratio for
the absorber.  Long and short term variability of the X--ray absorbing
column density would favour explanations based on an inner location for
the absorber (see e.g. Risaliti et al 2005 and also Lamer, Uttley \&
McHardy 2003). However, previous {\it ASCA} and {\it BeppoSAX}
observations show that the flux variability most likely occurs at
constant spectral slope and absorption.

The {\it XMM--Newton} data also reveal the presence (in both pn and
MOS detectors) of an absorption line with
rest--frame energy of $6.81^{+0.08}_{-0.06}$~keV, if the redshift of
IRAS~13197--1627 is assumed (z=0.01654). The absorption line energy
points towards an identification in terms of an Fe\textsc{xxv}
K$\alpha$ line which implies an outflow of $\sim 5000$~km~s$^{-1}$.
However, the outflow velocity is  consistent within the errors
with the systemic receding velocity of the galaxy (4959~km~s$^{-1}$)
raising the possibility that the gas responsible for the absorption is
instead local to our own Galaxy or its immediate surroundings.

We find that the most satisfactory description of the hard {\it
  XMM--Newton} X--ray spectrum can be obtained by considering an
additional reflection component from the accretion disc which
dominates the hard band and also accounts for the 15--30~keV Compton
hump seen in a previous {\it BeppoSAX} observation. The
reflection--dominated hard X--ray spectrum could be the sign that
strong gravitational effects are at work in IRAS~13197--1627 (see
Miniutti \& Fabian 2004).  Moreover, the disc reflection component is
correlated with the long--term continuum variability studied by using
also previous {\it ASCA} and {\it BeppoSAX} data over 10 years. This
behaviour is consistent with the light bending model which predicts a
good correlation between reflection and continuum only for
reflection--dominated sources, as it appears to be the case here but
the same prediction holds for any simple reflection model. The main
advantage of the light bending one is that the large observed
reflection fraction is automatically explained. It is worth mentioning
that, if the extreme observed reflection fraction ($R\sim 10$) is due
to light bending effects, the variability behaviour and the value of
$R$ itself imply that the primary AGN power law continuum is produced
within 4--5 gravitational radii from the black hole in a compact and
centrally concentrated active region (Miniutti \& Fabian 2004).

As mentioned by Dadina \& Cappi (2004) IRAS~13197--1627 shares a few
properties with the class of Narrow Line Seyfert~1 (NLS1) galaxies and
could represent their obscured counterpart. Here we also mention that
strong disc reflection components are more often seen (or claimed) in
NLS1 than in other type of X--ray sources and that the light bending
model seems to apply preferentially to NLS1s, thought to be
characterised by high mass--accretion rate (e.g. Fabian \& Miniutti
2006 for a review). This is not surprising given that disc reflection
is expected to occur preferentially in sources accreting in a
radiatively efficient manner, likely to be associated with a
relatively high accretion rate and with an accretion flow extending
down close to central black hole where most of the accretion power is
released. The steep hard X--ray spectra of NLS1s (and
IRAS~13197--1627) may indeed be an indication of a particularly high
accretion rate in this class (e.g. Shemmer et al 2006). As shown in
the next Section, a radiatively efficient accretion flow seems to be
indeed at work in IRAS~13197--1627. In the optical, narrow plus broad
Balmer lines are detected in IRAS~13197--1627 with the broad
components having a width of $\sim 1500$~km~s$^{-1}$, slightly wider
than the forbidden lines ($\sim 700$~km~s$^{-1}$), suggesting a
Seyfert 1.8 classification but leaving some room for a NLS1
interpretation (Aguero et al 1994). In this respect it is also very
interesting to consider that de Robertis et al (1988) reported a
number of low signal--to--noise broad features in the optical spectrum
which could be interpreted as {Fe\,\textsc{ii} emission, typical of
  NLS1 optical spectra.

  In the soft band below 2~keV the high--resolution RGS data indicate
  the presence of emission from photo--ionized gas in the form of N,
  O, and Ne emission lines as generally seen in absorbed type 2
  objects. However, the soft spectrum of IRAS~13197--1627 is more
  complex and we detect a plethora of emission lines due to a rich Fe
  L complex from Fe\textsc{xvi} and Fe\textsc{xvii}, that we interpret
  as due to thermal plasma emission.  Both the RGS and pn data can be
  described by considering a two--temperature thermal model in which
  the Fe L emission lines are due to plasmas at $\sim 0.6$~keV and
  $\sim 1.4$~keV. A power law--like continuum contributes by about 36
  per cent to the soft X--rays and could be due to i) scattering of
  the nuclear emission in the gas/dust revealed in emission/absorption
  or to ii) direct (and reflected) nuclear emission leaking through
  the absorber. The soft X--rays are instead dominated by the thermal
  components (50 per cent), while the photo--ionized gas emission
  represents about 14 per cent of the soft flux.

\subsection{Bolometric luminosity and Eddington ratio}

If the hard X--ray spectrum of IRAS~13197--1627 is indeed dominated by
disc reflection, computing the bolometric luminosity of the AGN from
the 2--10~keV one is a more difficult and ambiguous task than usual.
The bolometric luminosity of AGNs can obviously be computed from the
spectral energy distribution and usually compares well with the
8--1000$\mu m$ IR luminosity ($L_{\rm{IR}}$). In the present case,
$L_{\rm{IR}} \simeq 6.7\times 10^{44}$~erg~s$^{-1}$ (Sanders et al.
2003). However about half of it is likely to be associated with a
starburst (see discussion in next Section) and the AGN IR luminosity
can be estimated to be $L^{\rm{AGN}}_{\rm{IR}} \sim 3.1\times
10^{44}$~erg~s$^{-1}$.  $L^{\rm{AGN}}_{\rm{IR}}$ is most likely
produced by dust which intercepts the AGN emission and re--radiates it
in the IR. In our case, what is the X--ray luminosity that will be
intercepted by dust and re--radiated in the IR? This is not the
estimated AGN X--ray luminosity ($L^{\rm{est}}_{2-10}$, see Table~1
and relative discussion) because most of it is bent towards the inner
accretion disc and never escapes the immediate black hole
surroundings. What matters here is the (unabsorbed) observed
luminosity $L^{\rm{obs}}_{2-10}$, i.e. the hard X--ray luminosity that
can escape the gravitational sphere of influence of the black hole and
makes it to the outermost nuclear regions irradiating the dust
responsible for the IR re--emission.  Indeed, if one applies standard
bolometric corrections (Elvis et al 1994) to $L^{\rm{obs}}_{2-10}$
(see Table~1), one finds a bolometric luminosity of $\sim 2.1 \times
10^{44}$~erg~s$^{-1}$ which is in good agreement with $L_{\rm{IR}}$,
considering the likely large uncertainties in disentangling the
AGN--starburst contribution to $L_{\rm{IR}}$ (see next section).

However, if our spectral modelling and interpretation of the hard
X--ray spectrum are correct, the above estimate of the bolometric
luminosity has little to do with the real AGN power. Most of the power
is unobservable directly and is inferred because it produces the
strong reflection component. This means that $L^{\rm{est}}_{2-10}$ is
the quantity that represents the X--ray luminosity extracted from the
accretion process. Thus, by applying bolometric corrections to
$L^{\rm{est}}_{2-10}$, we can infer the ``true'' (or better the
``estimated'') bolometric luminosity of the AGN, i.e. the luminosity
that would be observed in absence of strong gravity and light bending
effects: for consistency, we shall call this quantity
$L^{\rm{est}}_{\rm{bol}}$. It turns out that
$L^{\rm{est}}_{\rm{bol}}\sim 1.3\times 10^{45}$~erg~s$^{-1}$. We
stress again that this is not the observed bolometric luminosity, but
represents instead the luminosity that would be observed f the primary
X--ray emission was isotropic and not gravitationally focused away
from the observer. Taking our interpretation to the extreme, we then
infer that, in absence of light bending effects, IRAS~13197--1627
would be seen as an ultra--luminous IR galaxy with $L_{\rm{IR}} \sim
10^{12} L_\odot$ because a much higher AGN luminosity would irradiate
the dust responsible for the AGN IR emission. 

$L^{\rm{est}}_{\rm{bol}}$ can also be used to compute the Eddington
ratio of IRAS~13197--1627 if the black hole mass can be estimated. To
our knowledge there is no black hole mass measurement for
IRAS~13197--1627. However, the velocity dispersion is $\sigma_* =
143$~km~s$^{-1}$ (Cid Fernandes et al.  2004) and, by using the
M$_{\rm BH}$--$\sigma_*$ relation (Gebhardt et al. 2000; Ferrarese \&
Merritt 2000) as derived by Greene \& Ho (2006) for local AGN, we
estimate a black hole mass of $1\div 3.5 \times 10^7~M_\odot$
corresponding to an Eddington luminosity L$_{\rm{Edd}} = 1.3\div
4.6\times 10^{45}$~erg~s$^{-1}$. If our interpretation is correct and
the AGN power is best estimated through $L^{\rm{est}}_{\rm bol}$,
IRAS~13197--1627 is radiating between 28 and 100 per cent of its
Eddington luminosity, although only a fraction of it is actually
observed at infinity and re--radiated e.g. in the IR. A high Eddington
ratio is consistent with the detection of a broad Fe line which
requires the accretion disc to extend down close to the black hole (as
likely in high Eddington rate sources), with the steep hard X--ray
spectrum which is a possible indicator of high mass--accretion rate
(Shemmer et al 2006), and with a NLS1 association (at least from the
X--ray point of view). For completeness, we also give the result based
on the observed 2--10~keV luminosity which is between 4.5 and 16 per
cent of the Eddington luminosity. All estimates given above should be
multiplied by about a factor 2 if the highest flux {\it BeppoSAX}
observation is considered.  As already mentioned by Dadina \& Cappi
(2004), the {\it BeppoSAX} data imply a 2--10~keV AGN estimated
luminosity of $\sim 10^{44}$~erg~s$^{-1}$, making IRAS~13197--1627 a
type 1.8 border--line Seyfert/quasar from the X--ray point of view.

\subsection{IRAS~13197--1627: a composite Seyfert/starburst galaxy?}

Besides emission from gas photo--ionized by the nucleus and
residual/scattered AGN emission, the soft X--ray spectrum is
characterised, and even dominated, by the presence of two thermal
components which could be the X--ray signature of star--formation. Cid
Fernandes et al (2004) estimate that a relatively young ($t<25$~Myr
and possibly $t\sim 5$~Myr) population of stars represent about the 25
per cent of the whole population in IRAS~13197--1627. The authors also
point out the detection of the 4680\AA \,\,bump associated with
Wolf--Rayet stars, an additional indication of star--formation
activity.

If the soft X--ray thermal component is indeed due to star--formation,
its luminosity in the 0.5--2~keV band ($1.4 \pm 0.4 \times
10^{41}$~erg~s$^{-1}$) can be used to infer the star--formation rate
(SFR) through the relation SFR$_{\rm X}=2.2\times
10^{-40}~L_{0.5-2~keV}~M_\odot$~yr$^{-1}$ (Ranalli, Comastri \& Setti
2003). We do not consider the luminosity of the photo--ionized gas
emission and continuum since these are due to AGN emission and we obtain
SFR$_{\rm X}\simeq 31\pm 9~M_\odot$~yr$^{-1}$. However, some
contribution from X--ray binaries may be masked in the soft X--ray
continuum: we assessed this possibility by forcing a $\Gamma\sim 1.2$
power law to be present in the X--ray spectrum (representing the
unresolved high mass X--ray binaries contribution, see e.g.  White et
al 1983; Persic et al 2004) obtaining an upper limit on any possible
additional contribution of $5~M_\odot$~yr$^{-1}$, i.e. SFR$_{\rm
  X}\simeq 31^{+14}_{-9}~M_\odot$~yr$^{-1}$.

On the other hand, IRAS~13197--1627 is a luminous infrared (LIR)
galaxy with $L_{\rm{8-1000~\mu m}} \equiv L_{\rm{IR}} = 6.7\times
10^{44}$~erg~s$^{-1}$ (Sanders et al. 2003). The IR luminosity is
generally considered as a good indicator of the star--formation rate
(SFR) in AGN--free galaxies, but the reprocessing of the AGN emission
in dust can dominate the IR luminosity.  Rowan--Robinson \& Crawford
(1989) deconvolved the IR spectral energy distribution of IRAS
galaxies into a disc, Seyfert, and starburst component finding that a
$\sim$45 per cent Seyfert and a $\sim$54 per cent starburst component
reproduce well the data of IRAS~13197--1627 (with only a marginal disc
contribution). Although the above numbers are probably affected by
large uncertainties, one can estimate a starburst IR luminosity of
$L^{\rm{SB}}_{\rm{IR}}\sim 3.6\times 10^{44}$~erg~s$^{-1}$. By assuming this
is correct, and by using SFR$_{\rm{IR}} = 4.5 \times 10^{44}
L^{\rm{SB}}_{\rm{IR}}~M_\odot$~yr$^{-1}$ (Kennicutt 1989), we derive
SFR$_{\rm{IR}} \sim 16~M_\odot$~yr$^{-1}$, slightly lower than
the SFR$_{\rm X}$ estimate given above.

As mentioned, IRAS~13197--1627 is also a radio source with a Radio
luminosity $L_{\rm{1.4~GHz}} \sim 1.6 \times
10^{30}$~erg~s$^{-1}$~Hz$^{-1}$ (Condon et al. 1996). Most of the
radio emission is probably due to the AGN and, if so, it is not a
useful indicator of SFR. At 2.3~GHz, the radio core contribution to
the total radio flux is about 15 per cent (Roy et al 1998). However,
even accounting for the radio core, IRAS~13197--1627 is still a member
of the ``radio--excess class'' among IRAS galaxies (Yun et al 2001),
meaning that its radio luminosity is in excess from that expected from
the well known tight radio--IR correlation for normal and starburst
galaxies. If we assume that IRAS~13197--1627 would lie on the
correlation if the AGN was removed, the fractional AGN contribution in
the radio band must be higher than its contribution in the IR. It
seams reasonable to assume that the AGN contributes by at least 50 per
cent at 1.4~GHz. If so, the AGN--free radio luminosity is
$L^{\rm{AGN-free}}_{\rm{1.4~GHz}} < 0.8 \times
10^{30}$~erg~s$^{-1}$~Hz$^{-1}$ and, by applying the relation SFR$_{\rm{1.4~GHz}} = 0.25 \times
10^{-28}~L^{\rm{AGN-free}}_{\rm{1.4~GHz}}$(e.g. Ranalli et al 2003) we thus obtain
SFR$_{\rm{1.4~GHz}} < 20~M_\odot$~yr$^{-1}$.

Summarising, we obtain SFR$_{\rm X}\simeq
31^{+14}_{-9}~M_\odot$~yr$^{-1}$,
SFR$_{\rm{IR}}\sim 16~M_\odot$~yr$^{-1}$, and SFR$_{\rm{1.4~GHz}} <
20~M_\odot$~yr$^{-1}$. At face value, the X--ray estimate seems to
overestimate the SFR obtained from IR and radio. This could be easily
explained if a fraction of the thermal emission we detect in the soft
X--rays was due to gas shocked by the radio jet and not to
star--forming regions. In the prototypical starburst galaxy M82, the
plasma is observed to have a multi--temperature distribution around
0.4--0.9~keV (Stevens, Read \& Bravo--Guerrero 2003), similar to the
lower--temperature component we detect in IRAS~13197--1627. It is thus
possible that the high--temperature $\sim$1.4~keV component we detect
is not associated with the starburst. In this respect, it is
interesting to note that if the high--temperature X--ray plasma is
associated with gas shocked by the radio jet and is excluded from the
SFR$_X$ estimate, we obtain SFR$_X \simeq 18^{+10}_{-5}$, more in line with
the IR and radio estimates.  A SFR of $15\div 20~M_\odot$~yr$^{-1}$
probably represents a reasonable value which can reconcile the
multi--wavelength data. The presence of a starburst could be the sign
that the nuclear region is gas rich, thus providing a large gas supply
for the accreting black hole which is consistent with the high
Eddington ratio inferred above.


\section*{Acknowledgements}

Based on observations obtained with XMM-Newton, an ESA science mission
with instruments and contributions directly funded by ESA Member
States and NASA. This research has made use of the Tartarus (Version
3.1) database, created by Paul O'Neill and Kirpal Nandra at Imperial
College London, and Jane Turner at NASA/GSFC. Tartarus is supported by
funding from PPARC, and NASA grants NAG5-7385 and NAG5-7067. G.
Miniutti thanks the PPARC for support. GP, MD, MC and G. Malaguti
thank ASI financial support under contract I/023/05/0.  GP also thanks
the European Commission under the Marie Curie Early Stage Research
Training Programme for supporting part of this research at the
Institute of Astronomy. G. Miniutti. thanks Kazushi Iwasawa, Stefano
Bianchi, Andy Fabian and Poshak Gandhi for useful discussions.We also would like
to thank the anonymous referee for a very useful report.

\end{document}